\documentclass[9pt]{elife}

\usepackage{subcaption}
\usepackage{cleveref}
\usepackage{placeins}
\newcommand{\suppfigref}[1]{Supplementary~\Cref{#1}}
\usepackage[flushleft]{threeparttable}

\graphicspath{{figures/}}
\usepackage{verbatim}
\usepackage[colorinlistoftodos]{todonotes}
\usepackage{xcolor}
\usepackage{pdfcomment}
\usepackage{etoolbox}
\usepackage[normalem]{ulem}
\usepackage{marginfix}

\newcommand{\papercommentmode}{pdf}
\newcommand{\commentstyle}[1]{%
  \ifstrequal{#1}{pdf}{\renewcommand{\papercommentmode}{pdf}}{%
    \ifstrequal{#1}{inline}{\renewcommand{\papercommentmode}{inline}}{%
      \ifstrequal{#1}{margin}{\renewcommand{\papercommentmode}{margin}}{%
        \ifstrequal{#1}{none}{\renewcommand{\papercommentmode}{none}}{%
          \PackageError{papercomments}{Unknown comment style '#1'}%
            {Use pdf, inline, margin, or none.}%
        }%
      }%
    }%
  }%
}

\DeclareRobustCommand{\papercomment}[4]{%
  \ifdefstring{\papercommentmode}{none}{#3}{%
  \ifdefstring{\papercommentmode}{pdf}{%
    \pdfmarkupcomment[markup=Highlight,color=#2]{#3}{#4}%
  }{%
    \ifdefstring{\papercommentmode}{margin}{%
      \uline{#3}%
      \ifcsdef{@captype}{%
        \todo[inline,color=#2!20]{\textbf{#1:} #4}%
      }{%
        \ifinner
          \todo[inline,color=#2!20]{\textbf{#1:} #4}%
        \else
          \todo[color=#2!20,linecolor=#2!50!black,size=\scriptsize]{\textbf{#1:} #4}%
        \fi
      }%
    }{%
      #3\todo[inline,color=#2!20]{\textbf{#1:} #4}%
    }%
  }%
  }%
}

\commentstyle{none} %

\title{Large Language Models Show Signs of Alignment with Human Neurocognition During Abstract Reasoning}

\author[1*]{Christopher Pinier}
\author[1]{Sonia Acuña Vargas}
\author[1]{Mariia Steeghs-Turchina}
\author[1]{Dora Matzke}
\author[1\authfn{1}]{Claire E. Stevenson}
\author[1\authfn{1}]{Michael D. Nunez}
\affil[1]{Psychological Methods, University of Amsterdam, Amsterdam, The Netherlands}

\corr{c.pinier@uva.nl}{CP}
\makeatletter
\renewcommand{\@correspondence}{%
  \href{mailto:c.pinier@uva.nl}{\nolinkurl{c.pinier@uva.nl}}\space(CP)}
\makeatother
\contrib[\authfn{1}]{These authors jointly supervised this work}

\begin{document}

\maketitle

\newcommand{\HierMatchedParticipants}{24}
\newcommand{\HierMatchedCompleteSessions}{96}
\newcommand{\HierPermutationCount}{10000}
\newcommand{\HierBootstrapCount}{10000}
\newcommand{\HierPermutationSeed}{0}
\newcommand{\HierBootstrapSeed}{1}
\newcommand{\HierUniquePatternPairs}{28}
\newcommand{\HierEEGMeasureCount}{3}
\newcommand{\HierChannelGroupCount}{5}
\newcommand{\HierPatternCount}{8}
\newcommand{\HierLLMCount}{8}
\newcommand{\HierNuisanceDiagnosticCount}{5}
\newcommand{\HierFRPParticipants}{24}
\newcommand{\HierFRPCompleteSessions}{96}
\newcommand{\HierResponseParticipants}{25}
\newcommand{\HierResponseSessions}{105}
\newcommand{\HierRestParticipants}{25}
\newcommand{\HierRestSessions}{105}
\newcommand{\HierFRPMeanRSA}{0.439}
\newcommand{\HierResponseMeanRSA}{0.192}
\newcommand{\HierRestMeanRSA}{0.053}
\newcommand{\HierResponseFrontalMeanRSA}{0.162}
\newcommand{\HierRestFrontalMeanRSA}{0.024}
\newcommand{\HierIndividualFRPMeanRSA}{0.146}
\newcommand{\HierIndividualResponseMeanRSA}{0.064}
\newcommand{\HierIndividualRestMeanRSA}{0.012}
\newcommand{\HierIndividualFrontalFRPMeanRSA}{0.188}
\newcommand{\HierIndividualFrontalResponseMeanRSA}{0.050}
\newcommand{\HierIndividualFrontalRestMeanRSA}{0.008}
\newcommand{\HierIndividualFrontalFRPRangeLow}{0.154}
\newcommand{\HierIndividualFrontalFRPRangeHigh}{0.208}
\newcommand{\HierIndividualFrontalFRPSDRangeLow}{0.24}
\newcommand{\HierIndividualFrontalFRPSDRangeHigh}{0.28}
\newcommand{\HierFullGroupFRPMeanRSA}{0.343}
\newcommand{\HierFullGroupResponseMeanRSA}{0.194}
\newcommand{\HierFullGroupRestMeanRSA}{0.201}
\newcommand{\HierFullGroupFrontalFRPMeanRSA}{0.359}
\newcommand{\HierFullGroupFrontalResponseMeanRSA}{0.186}
\newcommand{\HierFullGroupFrontalRestMeanRSA}{0.210}
\newcommand{\HierReliabilityFRPMeanRSA}{0.235}
\newcommand{\HierReliabilityFRPCILow}{0.153}
\newcommand{\HierReliabilityFRPCIHigh}{0.314}
\newcommand{\HierReliabilityResponseMeanRSA}{0.020}
\newcommand{\HierReliabilityResponseCILow}{-0.031}
\newcommand{\HierReliabilityResponseCIHigh}{0.070}
\newcommand{\HierReliabilityRestMeanRSA}{0.030}
\newcommand{\HierReliabilityRestCILow}{-0.018}
\newcommand{\HierReliabilityRestCIHigh}{0.078}
\newcommand{\HierReliabilityFrontalFRPMeanRSA}{0.238}
\newcommand{\HierReliabilityFrontalFRPCILow}{0.075}
\newcommand{\HierReliabilityFrontalFRPCIHigh}{0.388}
\newcommand{\HierReliabilityFrontalResponseMeanRSA}{-0.007}
\newcommand{\HierReliabilityFrontalResponseCILow}{-0.094}
\newcommand{\HierReliabilityFrontalResponseCIHigh}{0.081}
\newcommand{\HierReliabilityFrontalRestMeanRSA}{0.037}
\newcommand{\HierReliabilityFrontalRestCILow}{-0.035}
\newcommand{\HierReliabilityFrontalRestCIHigh}{0.108}
\newcommand{\HierFRPRegionalRangeLow}{0.270}
\newcommand{\HierFRPRegionalRangeHigh}{0.609}
\newcommand{\HierResponseRegionalRangeLow}{-0.018}
\newcommand{\HierResponseRegionalRangeHigh}{0.380}
\newcommand{\HierRestRegionalRangeLow}{-0.214}
\newcommand{\HierRestRegionalRangeHigh}{0.459}
\newcommand{\HierFRPModelRSARangeLow}{0.356}
\newcommand{\HierFRPModelRSARangeHigh}{0.500}
\newcommand{\HierNuisanceFRPFrontalFixationDurationRSA}{0.222}
\newcommand{\HierNuisanceFRPFrontalFixationDurationP}{0.003}
\newcommand{\HierNuisanceFRPOccipitalFixationDurationRSA}{0.328}
\newcommand{\HierNuisanceFRPOccipitalFixationDurationP}{<.001}
\newcommand{\HierNuisanceFRPFixationDurationRegionalRangeLow}{0.130}
\newcommand{\HierNuisanceFRPFixationDurationRegionalRangeHigh}{0.328}
\newcommand{\HierNuisanceFRPFrontalPositionRSA}{0.103}
\newcommand{\HierNuisanceFRPFrontalPositionP}{0.040}
\newcommand{\HierFRPResponseObservedZ}{0.281}
\newcommand{\HierFRPResponsePTwoSided}{0.276}
\newcommand{\HierFRPResponsePGreater}{0.139}
\newcommand{\HierFRPResponsePHolm}{0.276}
\newcommand{\HierFRPResponsePGreaterHolm}{0.139}
\newcommand{\HierFRPResponseCILow}{-0.076}
\newcommand{\HierFRPResponseCIHigh}{0.595}
\newcommand{\HierFRPRestObservedZ}{0.421}
\newcommand{\HierFRPRestPTwoSided}{0.088}
\newcommand{\HierFRPRestPGreater}{0.042}
\newcommand{\HierFRPRestPHolm}{0.175}
\newcommand{\HierFRPRestPGreaterHolm}{0.084}
\newcommand{\HierFRPRestCILow}{0.078}
\newcommand{\HierFRPRestCIHigh}{0.637}
\newcommand{\HierPatternLabelPermutationCount}{40320}
\newcommand{\HierPatternLabelFRPMeanRSA}{0.439}
\newcommand{\HierPatternLabelFRPPGreater}{0.040}
\newcommand{\HierPatternLabelFRPPHolm}{0.121}
\newcommand{\HierPatternLabelResponseMeanRSA}{0.192}
\newcommand{\HierPatternLabelResponsePGreater}{0.075}
\newcommand{\HierPatternLabelRestMeanRSA}{0.053}
\newcommand{\HierPatternLabelRestPGreater}{0.333}
\newcommand{\HierPatternLabelFrontalFRPObservedT}{0.710}
\newcommand{\HierPatternLabelOverallFRPObservedT}{0.481}
\newcommand{\HierPatternLabelOverallFRPPHolm}{0.121}
\newcommand{\HierPatternLabelFrontalResponseObservedT}{0.164}
\newcommand{\HierPatternLabelOverallResponseObservedT}{0.200}
\newcommand{\HierPatternLabelOverallResponsePHolm}{0.151}
\newcommand{\HierPatternLabelFrontalRestObservedT}{0.024}
\newcommand{\HierPatternLabelOverallRestObservedT}{0.060}
\newcommand{\HierPatternLabelOverallRestPHolm}{0.333}
\newcommand{\HierPatternLabelOverallFRPPGreater}{0.040}
\newcommand{\HierPatternLabelOverallResponsePGreater}{0.075}
\newcommand{\HierPatternLabelOverallRestPGreater}{0.333}
\newcommand{\HierPatternLabelFrontalFRPPGreater}{0.010}
\newcommand{\HierPatternLabelFrontalResponsePGreater}{0.226}
\newcommand{\HierPatternLabelFrontalRestPGreater}{0.453}
\newcommand{\HierPatternLabelFrontalFRPPHolm}{0.030}
\newcommand{\HierPatternLabelFrontalResponsePHolm}{0.451}
\newcommand{\HierPatternLabelFrontalRestPHolm}{0.453}
\newcommand{\HierFRPFrontalMeanRSA}{0.609}
\newcommand{\HierFRPTemporalMeanRSA}{0.270}
\newcommand{\HierFRPCentralMeanRSA}{0.404}
\newcommand{\HierFRPOccipitalMeanRSA}{0.469}
\newcommand{\HierFRPParietalMeanRSA}{0.444}

\newcommand{\BehParticipants}{25}
\newcommand{\BehSessions}{105}
\newcommand{\BehHumanAccuracyMean}{82.66}
\newcommand{\BehHumanAccuracySD}{19.83}
\newcommand{\BehModelAccuracyMean}{40.59}
\newcommand{\BehModelAccuracySD}{33.08}
\newcommand{\BehPeakTaskRSAMean}{0.71}
\newcommand{\BehPeakTaskRSASD}{0.03}
\newcommand{\BehAccuracyPeakCorrelation}{0.67}
\newcommand{\BehPeakDepthMean}{47.52}
\newcommand{\BehPeakDepthSD}{13.18}
\newcommand{\BehPhiProfileR}{0.78}
\newcommand{\BehPhiAccuracy}{32.00}
\newcommand{\BehGemmaTwoProfileR}{0.31}
\newcommand{\BehGemmaTwoAccuracy}{27.50}
\newcommand{\BehGemmaNineProfileR}{0.42}
\newcommand{\BehGemmaNineAccuracy}{8.50}
\newcommand{\BehGemmaTwentySevenProfileR}{0.22}
\newcommand{\BehGemmaTwentySevenAccuracy}{12.00}
\newcommand{\BehLlamaThreeProfileR}{0.02}
\newcommand{\BehLlamaThreeAccuracy}{7.50}
\newcommand{\BehLlamaSeventyProfileR}{0.42}
\newcommand{\BehLlamaSeventyAccuracy}{81.75}
\newcommand{\BehQwenProfileR}{0.75}
\newcommand{\BehQwenAccuracy}{80.50}
\newcommand{\BehDeepSeekProfileR}{0.75}
\newcommand{\BehDeepSeekAccuracy}{75.00}

\begin{abstract}
Large language models (LLMs) are increasingly approaching or surpassing human performance across a diverse array of tasks, thus raising a fundamental question: do their successes reflect representational structures similar to those found in humans? To investigate this, we compared \BehParticipants{} human participants and eight open-weight LLMs on an abstract-pattern-completion task, combining behavioral measures with representational similarity analysis of model activations and human fixation-related potentials (FRPs) recorded using electroencephalography (EEG). The three largest tested models approached human accuracy (\BehDeepSeekAccuracy{}--\BehLlamaSeventyAccuracy{}\% versus a human mean of \BehHumanAccuracyMean{}\%), but overall accuracy and similarity to the human pattern-specific difficulty profile yielded different model rankings. Beyond these behavioral differences, all eight models formed representations organized by abstract pattern types. This organization was most pronounced, on average, in intermediate layers. For each model, we selected the layer with maximal pattern-type distinction, independently of EEG, and compared its representations with human FRPs. This comparison showed that patterns represented similarly by the LLMs also tended to elicit similar frontal FRPs during sequence inspection (mean Pearson correlation across models, $r=\HierFRPFrontalMeanRSA{}$). This alignment was significant in an exact one-sided pattern-label permutation test ($p=\HierPatternLabelFrontalFRPPGreater{}$), supporting our primary hypothesis of shared representational structure. The corresponding comparisons with frontal response-locked and pre-trial rest EEG yielded lower mean alignment. These findings identify correspondence between model representations and human frontal EEG during abstract-pattern completion, without establishing that humans and LLMs use the same reasoning mechanisms.

\end{abstract}

\section{Introduction}
The prospect of developing an intelligent system akin to human intelligence has long captivated the public imagination. Once relegated to science fiction and thought experiments, the idea has entered the realm of possibilities ever since the appearance of the first computer systems. Successive breakthroughs, from computers mastering Chess and Go \citep{silver_mastering_2016, silver_mastering_2017, silver_general_2018} to the public release of ChatGPT, have steadily expanded the range of tasks at which machines rival or surpass human performance. Large language models (LLMs) now provide a single architecture that approaches human‑level competence across diverse linguistic and reasoning challenges. Yet whether these models reason in a genuinely human‑like manner remains an open question. One way to probe this question is through \emph{abstract reasoning}, a fundamental ability to extract patterns, rules, and relationships from limited information and apply them in new contexts. Reasoning about abstract relations is central to fluid intelligence and supports flexible problem solving \citep{ferrer_fluid_2009, chuderski_fluid_2022}, and the aspiration to build artificial systems with these abilities dates back to the earliest days of AI research. Newell and Simon’s pioneering chess program and logic theory machine framed intelligent behavior as heuristic search over symbolic abstractions \citep{newell_chess_1955, newell_logic_1956}. Now, 70 years later, LLMs appear closer to that goal, solving numerous abstract‑reasoning tasks with human‑level proficiency \citep{bubeck_sparks_2023, musker_llms_2025, webb_emergent_2023, webb_evidence_2025}. However, if LLMs' reasoning is truly human‑like, we might not only expect human‑like responses but also human‑like internal representations and computational processes. To investigate this, we extend recent work aligning human neural representations with LLM representations on perceptual and linguistic tasks \citep[e.g., ][]{doerig_visual_2024, lei_large_2025} to the realm of abstract reasoning and compare human performance and neural representations to those of eight open-weight LLMs while solving an abstract-pattern-completion task.

\subsection*{LLMs' alignment with human behavior and cortical patterns}
    The early and mid-2010s saw the rise of a remarkable, albeit narrow, form of artificial intelligence (AI), with deep neural networks achieving near-human accuracy in domain-specific tasks such as image classification and object recognition \citep[e.g., ][]{krizhevsky_imagenet_2012, lecun_deep_2015}. Despite their success, they offered limited insight into general cognitive mechanisms and often failed to generalize beyond their training distributions \citep[for a review on their limitations, see ][]{bowers_deep_2022}. The arrival of transformer-based LLMs marked a profound shift, both in the field of AI and in the cognitive sciences. Trained predominantly on vast text corpora, these models have demonstrated a broad spectrum of capabilities, matching or exceeding human performance on selected benchmarks in graduate-level science, multimodal reasoning, and competition mathematics \citep{sajadieh_ai_index_2026,balunovic_matharena_2025}. This rapid progress, coupled with the increasing availability of open-weight LLMs, has opened up exciting research avenues enabling neuroscience-like experiments on the internal activations of these models.  %

    A growing body of work shows that the latent spaces of LLMs (the representations that emerge within their hidden layers that capture some underlying features of their training data) mirror the implicit representational spaces that guide human behavior. \citet{dima_shared_2024}, for instance, found that text vectors from a Generative Pre-trained Transformer (GPT) embedding model (text-embedding-ada-002) explained participants’ perceived similarity of human actions in both naturalistic sentences and videos better than other language and vision models. \citet{iaia_representational_2025} also found a significant alignment between the implicit representational geometry of humans and Bidirectional Encoder Representations from Transformers (BERT), among other language models, on a semantic odd-one-out task.
    \citet{marjieh_large_2024} corroborates this trend, this time at the perceptual level, showing that similarity judgements from GPT models (versions 3 to 4) accurately reproduce the structured ways humans perceive and organize sensory information, such as the color wheel and the pitch spiral. %
    These parallels naturally prompt a deeper question: do the same internal states that let LLMs mirror human psychology also match cortical patterns? Converging evidence suggests they do, with studies showing that these internal states seem to be systematically predictive of activity in the human cortex during linguistic \citep{caucheteux_brains_2022, goldstein_alignment_2024, lei_large_2025, mischler_contextual_2024, schrimpf_neural_2021} or visual \citep{doerig_visual_2024} tasks. \citet{schrimpf_neural_2021} showed that GPT-2 best predicted human brain responses on naturalistic reading and listening tasks compared to 43 language models drawn from multiple neural network families, achieving near-ceiling fits across three complementary datasets, involving functional magnetic resonance imaging (fMRI) and electrocorticography (ECoG) recordings. Extending this line of work, \citet{lei_large_2025} investigated LLM–brain predictivity  during naturalistic story listening more closely by comparing layer-wise activations from fourteen LLMs, in both their base and instruction-tuned versions, with fMRI recordings. They found that the instruction-tuned versions consistently outperformed their base counterparts and that higher performance was positively associated with brain correlations. Additionally, in accordance with previous findings, they found that brain-predictive power peaked in the LLMs’ intermediate layers \citep{caucheteux_brains_2022, mischler_contextual_2024}.

    In summary, the collective evidence now spans multiple independent datasets and neural modalities, including fMRI, electroencephalography (EEG), and ECoG, painting an increasingly coherent picture: when LLMs achieve human-like behavioral competence, their internal activation spaces tend to mirror the brain’s representational landscape along multiple spatial and temporal scales. However, to date, most studies have demonstrated alignment on perceptual or linguistic tasks where low-level features, such as token frequency or sentence length, can spuriously inflate brain-score metrics  \citep{feghhi_what_2024}. Because abstract reasoning demands rule induction and relational generalisation that go beyond such surface statistics, it offers a more stringent benchmark.
    
\subsection*{Investigating reasoning processes between humans and AI via abstractions}
    As described earlier, abstract reasoning corresponds to the ability to identify patterns, rules, and relationships from limited information and apply them to new or different contexts. This capacity for higher-order thinking is tightly linked to the concept of fluid intelligence \citep{ferrer_fluid_2009, chuderski_fluid_2022}, which is thought to be supported by cortical networks in frontal and parietal brain regions \citep{caudle_neural_2023, choi_multiple_2008, duncan_multiple-demand_2010, gray_neural_2003, perfetti_differential_2007, santarnecchi_dissecting_2017, tschentscher_fluid_2017, zurrin_functional_2024}. %
    Within these networks, lesion evidence implicates the lateral frontal cortex in generating abstract rules, while functional-imaging studies associate frontal activity with discovering abstract action rules and integrating relational information \citep{reverberi_generation_2005, badre_frontal_2010, dumontheil_development_2014, davis_concrete_2017}.
    Abstract reasoning therefore provides a useful domain for examining both behavioral generalization and representational correspondence between humans and artificial systems.

    Recent advances in LLMs have substantially improved performance on reasoning tasks, supported by reinforcement-learning post-training and additional computation at inference time. Reviews of these developments describe improvements in generating, exploring, and evaluating candidate solutions, including methods specifically targeting inductive reasoning---the inference of general rules from particular examples \citep{xu_large_2025,chen_survey_2026}. These advances motivate increasingly precise comparisons with human cognition: how reliably do models generalize an inferred rule across changes in task presentation, and to what extent do their solutions depend on representations resembling those used by humans?

    Evidence from analogical reasoning illustrates why these questions require attention to both task conditions and model capabilities. \citet{lewis_evaluating_2024} found that the GPT models they tested were more sensitive than humans to several modifications of analogy problems, including changes to letter sequences and answer ordering. However, subsequent experiments showed that GPT-4 augmented with code execution could achieve approximately human-level performance on counterfactual letter-string analogies involving a permuted alphabet. Additional controls implicated difficulties with counting and indexing as contributors to earlier failures \citep{webb_evidence_2025}. 

    Direct analyses of internal representations offer a complementary approach. For example, mechanistic work has identified attention-head computations supporting symbol abstraction and induction over abstract sequences in LLMs \citep{yang_emergent_2025}. Whether such representations correspond to those formed by humans remains a separate empirical question. Here, we investigate this question by comparing the representational geometry of abstract patterns in model activations and human EEG recorded during corresponding pattern-completion tasks. This comparison tests whether patterns represented similarly by models also elicit similar human neural responses. Such correspondence would provide evidence of shared representational structure, without requiring identical computational mechanisms or establishing that models and humans use the same solution strategies.

    \subsection*{Using fixation-related potentials as a temporally-precise window into human cognition}

    Traditional experimental paradigms often rely on presenting stimuli sequentially while participants maintain static fixation and provide minimal responses, such as button presses. Researchers inspecting EEG and fMRI data collected during these experimental paradigms then lock neural measurements to experimenter-defined events (e.g., stimulus onset, button press). But these methods risk missing the strategy employed by participants during reasoning because of trial-to-trial variability in neurocognitive processing \citep[e.g. see ][]{weindel_trial-by-trial_2024,weindel_decision-making_2025}.

    One way to overcome these limitations is to combine brain activity recording with eye-tracking. Eye-tracking provides a high-resolution record of measures thought to reflect overt visual attention, such as where a person is looking (fixations) and how they move their eyes (saccades), while EEG captures the brain’s electrical activity with millisecond precision. By synchronizing these two data streams, it becomes possible to analyze neural activity time-locked to specific oculomotor events that occur during natural viewing behavior.

    \emph{Fixation-related potentials} (FRPs) are derived from this co-registration approach. They represent averaged segments of the ongoing EEG signal aligned to the onset of gaze fixations. In essence, FRPs are analogous to the more conventional visual stimulus-locked event-related potential (ERP) but different because eye movements are self-driven and reflect ongoing cognitive processing \citep[for a review, see][]{degno_eye_2020}. We used FRPs to examine neural activity while participants inspected the abstract sequences. Eye-movement patterns have been linked to differences in solution strategy during visual reasoning tasks \citep{kucharsky_cognitive_2020,liu_exploring_2025}, motivating the use of fixation onset to sample neural activity during participants' ongoing inspection of the sequence. We therefore compared the pattern-level representational geometry of FRPs with that of LLM activations, using response-locked and pre-trial rest EEG as comparison measures.

    \subsection*{Current study}
\begin{figure}[!htbp]
    \centering
    \includegraphics[width=\linewidth]{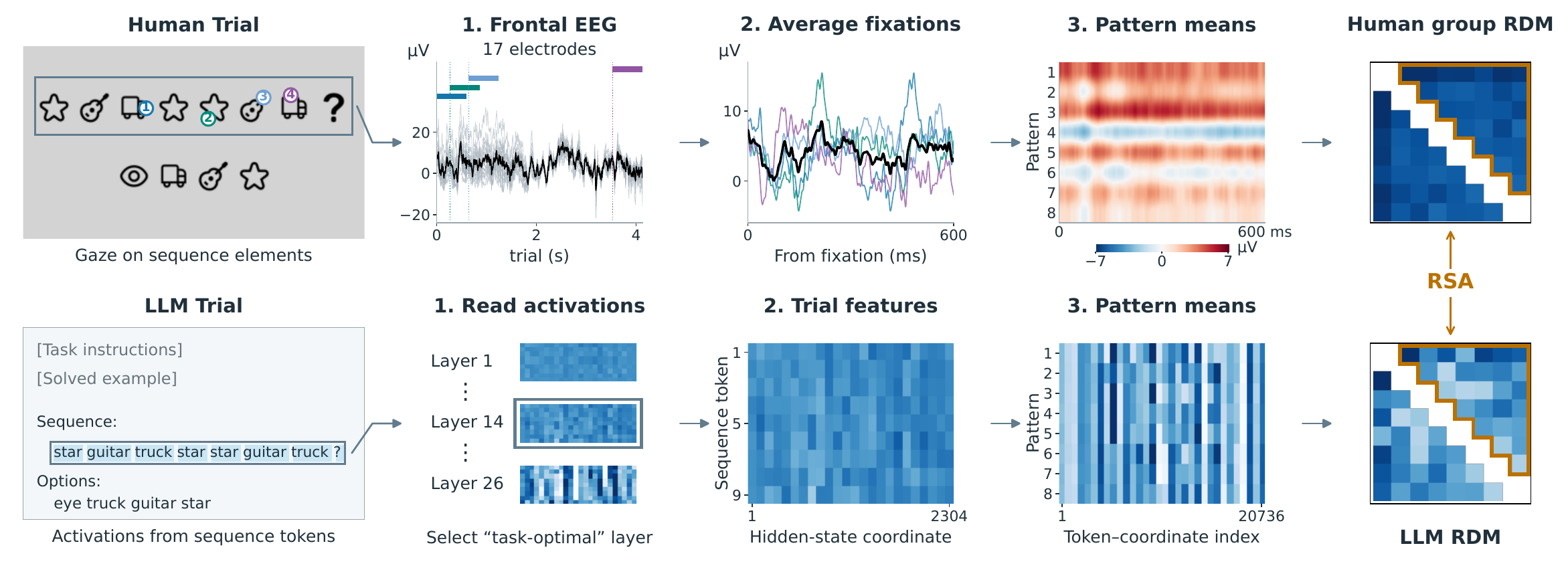}
    \caption{Experimental task and construction of human and LLM representations. Both rows illustrate the same task item. \textbf{Human processing (top):} Numbered gaze markers identify four retained sequence fixations. Panel 1 shows 17 frontal-electrode traces (gray), their mean (black), and colored 600-ms fixation-locked extraction windows; time is relative to the first retained fixation. Panel 2 averages the corresponding epochs (colored) into a trial FRP (black). Panel 3 shows eight pattern-mean FRPs from the example participant and session; red and blue indicate positive and negative voltages. Session RDMs are averaged within participants, then equally across participants. \textbf{LLM processing (bottom):} Blue highlights identify sequence words used for activation extraction. Panel 1 shows Gemma-2-2B layers 1, 14, and 26; layer 14 is task-optimal, selected independently of EEG (see \hyperref[meth:llm-rdms]{Methods, \textit{LLM RDMs}}). Its token activations (panel 2) are concatenated into trial vectors and averaged within patterns (panel 3). Activation displays show a subset of coordinates; RDMs use the full vectors. \textbf{RSA} correlates the 28 corresponding upper-triangle entries, outlined in orange. RDM color scales are independent.}
    \label{fig:experiment-overview}
\end{figure}

    This study is structured around two objectives: i) to assess whether current open-weight LLMs can perform a relatively simple reasoning task involving arbitrary symbols and abstract patterns in a manner that exhibits similar behavioral patterns to humans, ii) to explore whether the internal representations formed within these LLMs align with human cortical activity recorded via EEG. In our task, 25 human participants solved abstract-pattern-completion problems while EEG signals and gaze data were simultaneously recorded. Each trial consisted of a sequence of icons spatially arranged according to an implicit logical rule, such as ABCDDCBA or ABBACDDC, that we refer to as a ``pattern'' (see the example trial at the upper left of \Cref{fig:experiment-overview}). Eight open-weight LLMs were presented with a text-based version of the task, replacing the icons with corresponding one-word labels. The performance and hidden-layer activations of these LLMs were then compared to human behavior and neural activations.

    To compare human and model representations, we used representational similarity analysis (RSA). For each modality, a representational dissimilarity matrix (RDM) summarized how differently the eight abstract patterns were represented. Correlating corresponding entries in the human and model RDMs allowed us to ask whether patterns represented similarly by an LLM also elicited similar human EEG responses.

    Based on evidence implicating lateral frontal cortex in the generation and discovery of abstract rules \citep{reverberi_generation_2005,badre_frontal_2010}, our primary neural hypothesis was that the representational geometry of abstract patterns in LLM activations would correspond to that of human FRPs measured over frontal electrodes during sequence inspection. Response-locked and pre-trial rest EEG served as secondary control measures. We analyzed EEG representational geometry across five non-overlapping electrode groups (frontal, central, temporal, parietal, and occipital). To our knowledge, this is the first study to test whether LLM and human EEG pattern geometries exhibit measurable representational similarity during \textit{abstract reasoning}.

\FloatBarrier
\section{Results}
\subsection*{Behavioral performance and abstract-pattern representations}

\begin{figure}[!htbp]
    \centering
    \includegraphics[width=\linewidth]{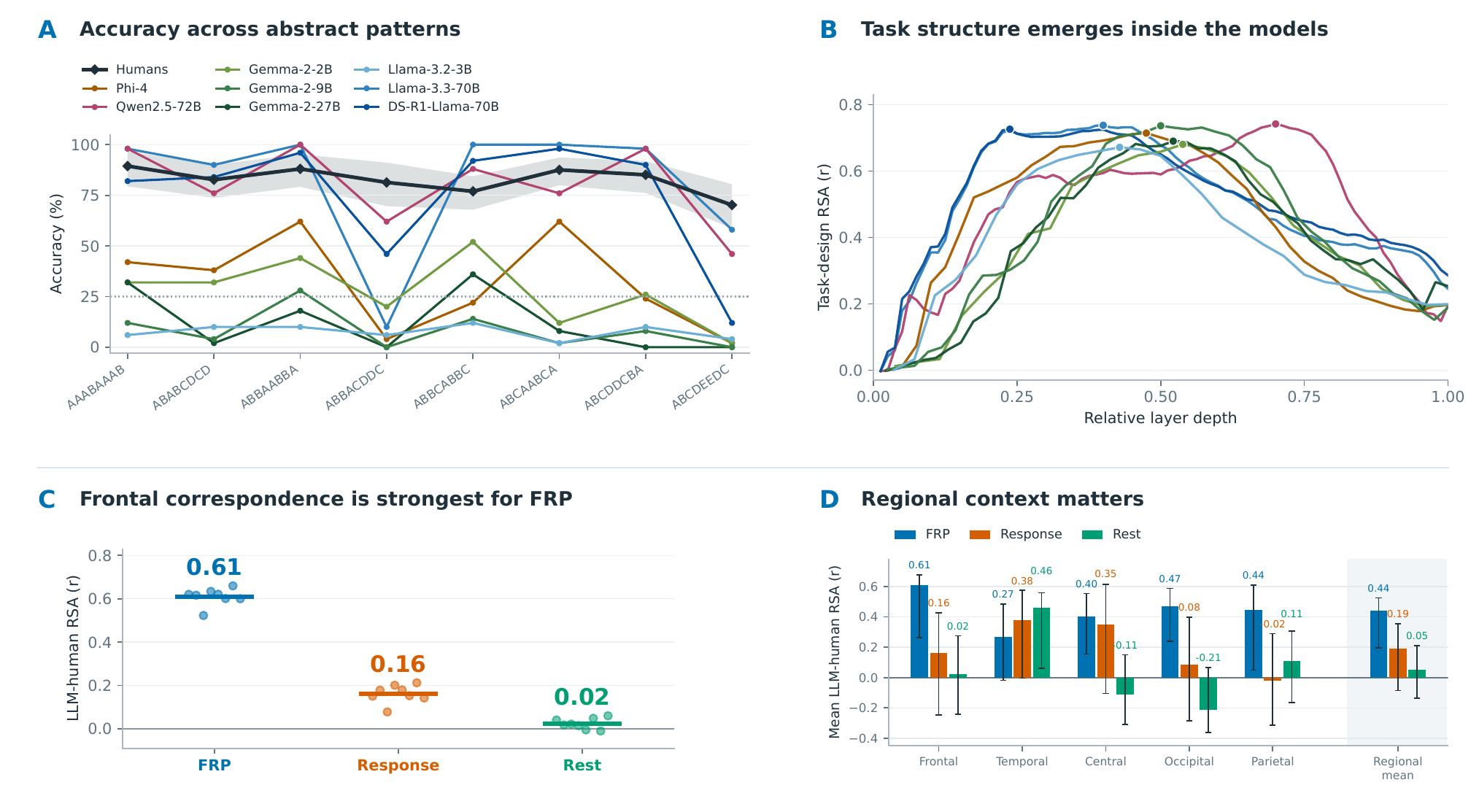}
    \caption{Main behavioral and representational results. \textbf{A:} Accuracy across eight abstract patterns. The human curve shows equally weighted participant means from \BehSessions{} sessions shared with the response/rest EEG exports ($N=\BehParticipants{}$); shading denotes pointwise 95\% bootstrap confidence intervals. The dotted line marks 25\% chance accuracy. \textbf{B:} Layer-wise correspondence with the task-design RDM; dots mark each model's maximum. \textbf{C:} Frontal LLM--human RSA for FRP, response-locked, and pre-trial rest EEG. Dots represent the eight LLMs; horizontal marks and labels show arithmetic means. \textbf{D:} Mean RSA across models for each electrode group; the shaded final group averages the five regional means. Error bars show percentile 95\% confidence intervals from 10,000 paired participant-bootstrap resamples, rebuilding the human group RDM while holding the eight LLMs fixed.
    Panels C--D use the same matched sample of \HierMatchedParticipants{} participants and \HierMatchedCompleteSessions{} sessions.
    }
    \label{fig:main-results}
\end{figure}

\begin{figure}[!htbp]
    \centering
    \includegraphics[width=0.65\linewidth]{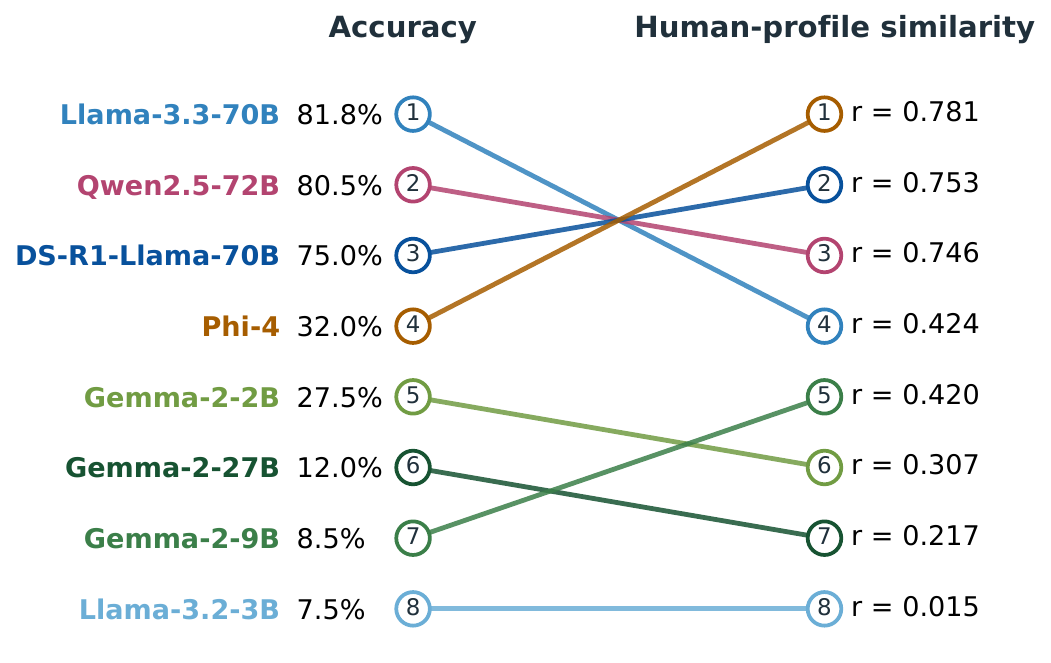}
    \caption{LLM rankings by accuracy and similarity to the human difficulty profile. Lines connect each model's rank in overall task accuracy (left) to its rank in Pearson correlation with the human accuracy profile across the eight abstract patterns (right). Circled numbers on both sides indicate ranks, with 1 highest. Percentages on the left report overall accuracy; $r$ values on the right are Pearson correlations between model and human mean accuracies across the eight patterns, with higher values indicating more similar difficulty profiles. Vertical positions represent ranks rather than numerical distances. Related model families use similar colors, with blue shades for Llama models and the Llama-distilled DeepSeek model, and green shades for Gemma models. The human profile gives each participant equal weight within each pattern; ranks use unrounded correlations.}
    \label{fig:accuracy-profile-ranks}
\end{figure}

\noindent\textbf{Accuracy and human-like difficulty profiles are distinct.} We defined a pattern-specific difficulty profile as the vector of accuracies across the eight abstract patterns. Human participants ($N=\BehParticipants{}$) outperformed the eight LLMs on average (human accuracy: $M=\BehHumanAccuracyMean{}\%, SD=\BehHumanAccuracySD{}\%$; LLM accuracy: $M=\BehModelAccuracyMean{}\%, SD=\BehModelAccuracySD{}\%$), although the three largest models ($\sim$70-billion parameters) had accuracies ranging from \BehDeepSeekAccuracy{}\% to \BehLlamaSeventyAccuracy{}\% (\Cref{fig:main-results}A; \Cref{fig:accuracy-profile-ranks}). The ABCDEEDC pattern was the most difficult for both groups. Qwen2.5-72B and DeepSeek-R1-Distill-Llama-70B (DS-R1-Llama-70B) combined high accuracy with relatively human-like pattern-specific difficulty profiles (both reaching $r=\BehQwenProfileR{}$), whereas Llama-3.3-70B was highly accurate but less similar to the human profile ($r=\BehLlamaSeventyProfileR{}$). Phi-4 showed the opposite relationship, with a low task accuracy (\BehPhiAccuracy{}\%) but a relatively human-like difficulty profile ($r=\BehPhiProfileR{}$). High accuracy therefore did not consistently imply human-like differences in difficulty across patterns.

\begin{figure}[!htbp]
    \centering
    \includegraphics[width=0.9\linewidth]{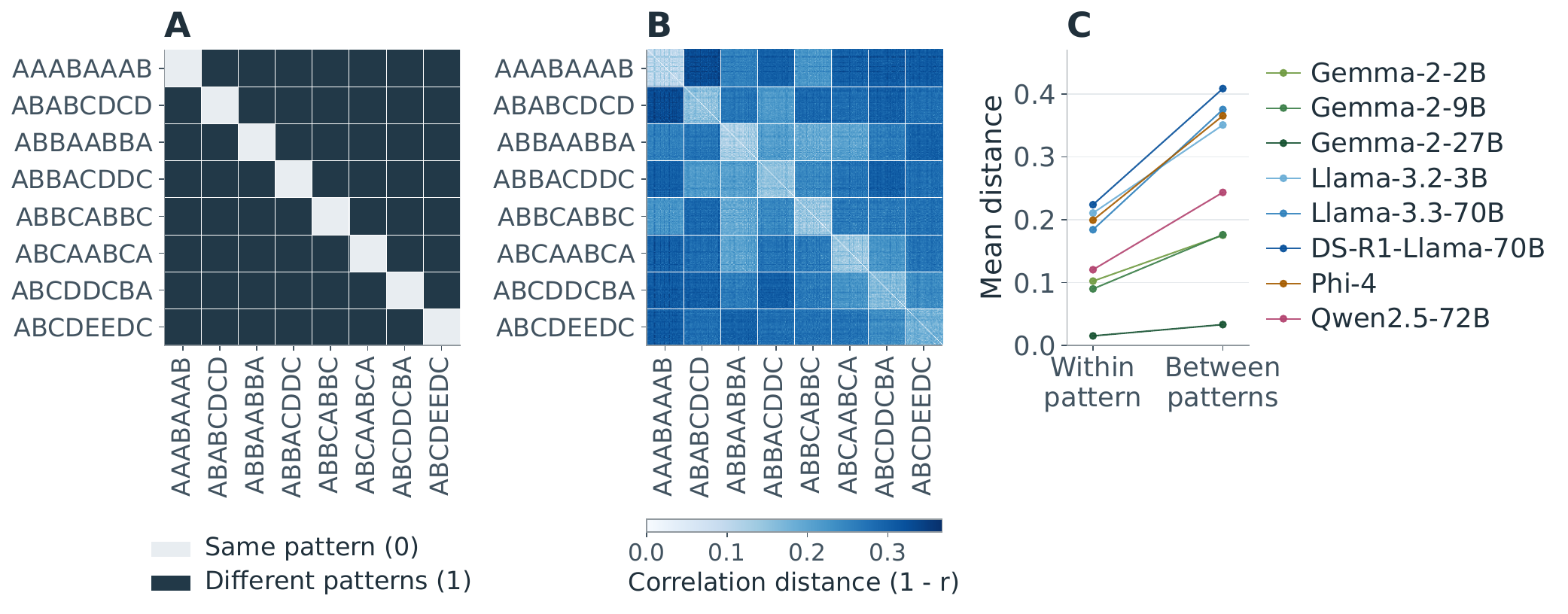}
    \caption{Trial-level representational dissimilarity matrices and within-/between-pattern distances. \textbf{A:} Task-design reference RDM, assigning zero dissimilarity to trials from the same abstract pattern and unit dissimilarity to trials from different patterns. \textbf{B:} Mean of the eight LLM trial RDMs, each computed at that model's task-optimal layer. Trials are grouped by abstract pattern in the same order in A and B. Lighter colors indicate lower dissimilarity in both panels, although their numerical scales differ. \textbf{C:} Each colored line connects one model's mean within-pattern and between-pattern correlation distances at its task-optimal layer. Within-pattern means exclude self-distances.}
    \label{fig:reference-and-llm-rdms}
\end{figure}

\noindent\textbf{Abstract-pattern structure is strongest in intermediate model layers.} We compared each layer's trial RDM with a $400\times400$ task-design RDM that distinguished within-pattern from between-pattern trial pairs, independently of human EEG. For each model, the \textit{task-optimal layer} was the layer whose trial-level RDM had the highest correlation with the task-design RDM. Across models, similarity rose from near zero in early layers, peaked in intermediate blocks (mean depth $\BehPeakDepthMean{}\%$, $SD=\BehPeakDepthSD{}\%$), and declined toward the output layer (\Cref{fig:main-results}B; \suppfigref{fig:llm-task-layer-profiles}). At each model's task-optimal layer, representations of trials sharing the same abstract pattern were more similar, on average, than representations of trials with different patterns (\Cref{fig:reference-and-llm-rdms}C). Across models, peak similarity to the task-design RDM averaged $\BehPeakTaskRSAMean{}$ ($SD=\BehPeakTaskRSASD{}$). Models with higher peak similarity also tended to achieve higher task accuracy (Pearson's $r=\BehAccuracyPeakCorrelation{}$, across eight models). Selecting the peak layer for each model and averaging its trial vectors within pattern yielded the eight-pattern LLM RDMs used in the neural comparison.

\subsection*{Frontal FRPs show pattern-specific alignment with LLM representations}

We compared the LLM RDMs with participant-averaged human RDMs from \HierMatchedCompleteSessions{} sessions contributed by \HierMatchedParticipants{} participants. One of the \BehParticipants{} participants had no session with complete FRP coverage across all eight patterns and five electrode groups (see \hyperref[sec:sample-inclusion]{Methods, \textit{Sample inclusion}}). Each RSA value corresponds to the Pearson correlation across the \HierUniquePatternPairs{} unique pattern pairs. The task-optimal LLM RDMs and the human frontal-FRP group RDM are shown in \suppfigref{fig:supp-optimal-layer-rdms}. Frontal FRP alignment was the single primary neural hypothesis. Response-locked and pre-trial rest EEG served as control measures, and the other four electrode groups provided spatial context (\suppfigref{fig:eeg-channel-groups}).

\begin{figure}[!htbp]
    \centering
    \includegraphics[width=0.90\linewidth]{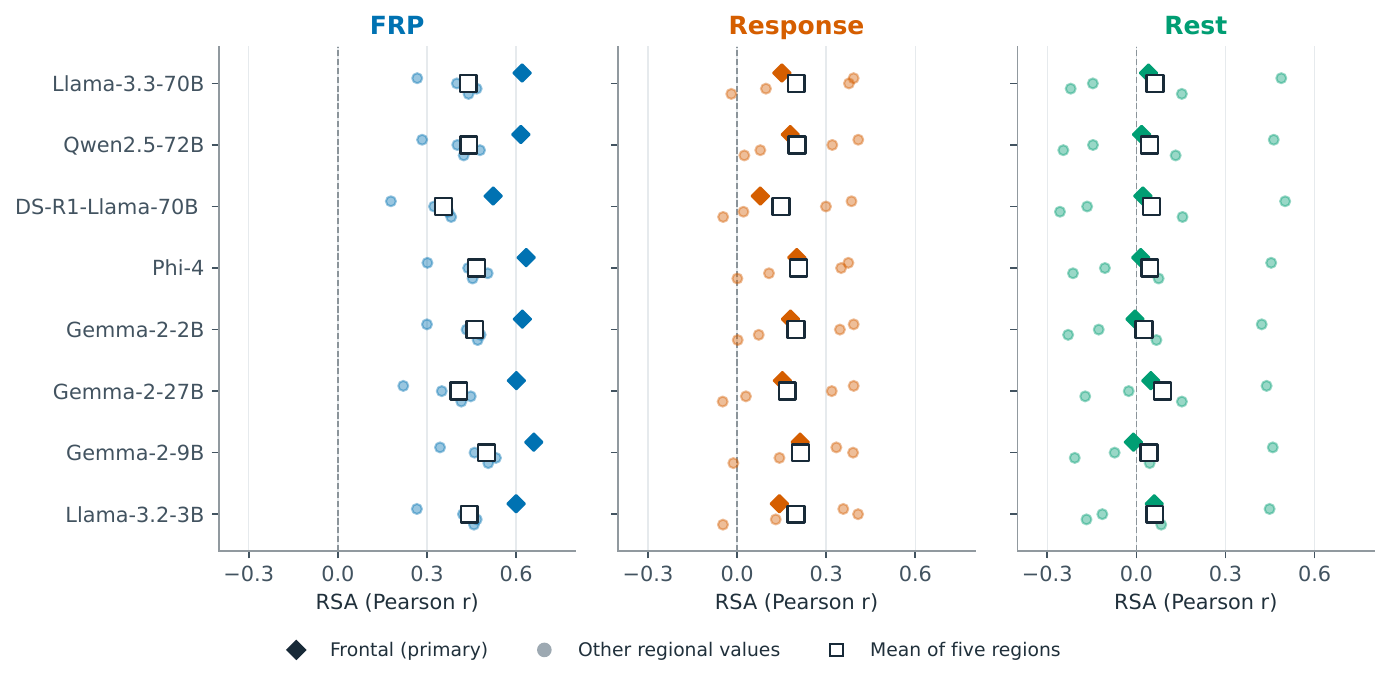}
    \caption{Model-specific LLM--human representational similarity across EEG measures and regions. All comparisons use the same matched set of \HierMatchedParticipants{} participants and \HierMatchedCompleteSessions{} complete participant--session cells. LLM layers were selected using the task-design RDM, independently of human EEG. Each model is shown separately for FRP (blue), response-locked EEG (orange), and pre-trial rest EEG (green). Diamonds identify frontal channels, the primary regional target; circles show the other regional values, and open squares show the mean across regions. Models are ordered by task accuracy. Regional points describe spatial variation and are not confidence intervals. DS-R1-Llama-70B abbreviates DeepSeek-R1-Distill-Llama-70B.}
    \label{fig:llm-group-rsa-by-model}
    \label{fig:matched-llm-human-rsa}
\end{figure}

\noindent\textbf{Group-level correspondence is strongest for frontal FRPs.} Mean frontal RSA was \HierFRPFrontalMeanRSA{} for FRP, compared with \HierResponseFrontalMeanRSA{} for response-locked EEG and \HierRestFrontalMeanRSA{} for pre-trial rest EEG. FRP correspondence was positive in all five regions (regional means \HierFRPRegionalRangeLow{}--\HierFRPRegionalRangeHigh{}), with across-region means of \HierFRPMeanRSA{}, \HierResponseMeanRSA{}, and \HierRestMeanRSA{} for FRP, response, and rest, respectively (\Cref{fig:main-results}C--D). Every LLM had positive FRP RSA when averaged across regions (\HierFRPModelRSARangeLow{}--\HierFRPModelRSARangeHigh{}), but their ordering differed from the behavioral ranking. For example, the two most accurate models, Llama-3.3-70B and Qwen2.5-72B, were not the models with the highest mean FRP RSA across regions (\Cref{fig:llm-group-rsa-by-model}).

\noindent\textbf{Frontal FRPs show pattern-specific alignment.} Holding the LLM RDMs fixed, we enumerated all $\HierPatternCount{}!=\HierPatternLabelPermutationCount{}$ joint row-and-column relabellings of the human RDMs. The test statistic, $T$, was the mean Fisher-transformed RSA across the eight models. Consistent with our primary hypothesis, frontal FRP alignment was significant in this pattern-label permutation test at $\alpha=0.05$ ($T=\HierPatternLabelFrontalFRPObservedT{}$, exact one-sided permutation $p=\HierPatternLabelFrontalFRPPGreater{}$). The same tests for the frontal response-locked and pre-trial rest controls were not significant ($p=\HierPatternLabelFrontalResponsePGreater{}$ and $p=\HierPatternLabelFrontalRestPGreater{}$, respectively). The frontal FRP finding also remained significant under the supplementary, broader Holm correction across all three frontal measures (adjusted $p=\HierPatternLabelFrontalFRPPHolm{}$). Full corrected frontal and exploratory across-region results are reported in the supplement (\hyperref[sec:supp-pattern-tests]{Supplementary results: exact pattern-identity alignment}). Thus, the primary test supports frontal FRP alignment tied to the intended pattern identities.

\noindent\textbf{Frontal FRP correspondence extends beyond the task-optimal layer.} While task-design RSA peaked in the middle layers (\Cref{fig:main-results}B), frontal FRP RSA extended across middle and later layers (\suppfigref{fig:supp-layerwise-eeg-regions}A). Frontal response-locked RSA was lower over much of this range, and frontal rest RSA was near zero (\suppfigref{fig:supp-layerwise-eeg-regions}B--C). Rest RSA remained positive in temporal and negative in occipital channels across middle and later layers.

\subsection*{Shared human geometry coexists with individual variability}

\noindent\textbf{FRPs show the highest participant-to-group similarity.} To assess the participant-shared structure underlying the group-level comparison, we correlated each participant's RDM with a group RDM formed from all other participants. Across regions, mean leave-one-out similarity was $r=\HierReliabilityFRPMeanRSA{}$ (95\% confidence interval (CI) [\HierReliabilityFRPCILow{}, \HierReliabilityFRPCIHigh{}]) for FRP, compared with \HierReliabilityResponseMeanRSA{} for response and \HierReliabilityRestMeanRSA{} for rest. Frontal channels showed the same ordering: \HierReliabilityFrontalFRPMeanRSA{}, \HierReliabilityFrontalResponseMeanRSA{}, and \HierReliabilityFrontalRestMeanRSA{}, respectively (\suppfigref{fig:participant-human-reliability}; full sample sizes and intervals in \hyperref[sec:supp-human-reliability]{Supplementary results: human reliability}). FRP geometry thus contained a shared component alongside substantial individual variability.

\noindent\textbf{LLM alignment is weaker at the individual-participant level.} We next examined how the group-level LLM--human correspondence compared with alignment measured separately in individual participants. Mean individual-participant frontal RSA was \HierIndividualFrontalFRPMeanRSA{} for FRP, \HierIndividualFrontalResponseMeanRSA{} for response, and \HierIndividualFrontalRestMeanRSA{} for rest, compared with the corresponding group-level values of \HierFRPFrontalMeanRSA{}, \HierResponseFrontalMeanRSA{}, and \HierRestFrontalMeanRSA{}. For frontal FRPs, model-specific participant means ranged from \HierIndividualFrontalFRPRangeLow{} to \HierIndividualFrontalFRPRangeHigh{}, with standard deviations of \HierIndividualFrontalFRPSDRangeLow{}--\HierIndividualFrontalFRPSDRangeHigh{} (\hyperref[sec:supp-individual-rsa]{Supplementary results: individual-participant RSA}).

\subsection*{Fixation duration covaries with FRP geometry}

\begin{figure}[!htbp]
    \centering
    \includegraphics[width=1\linewidth]{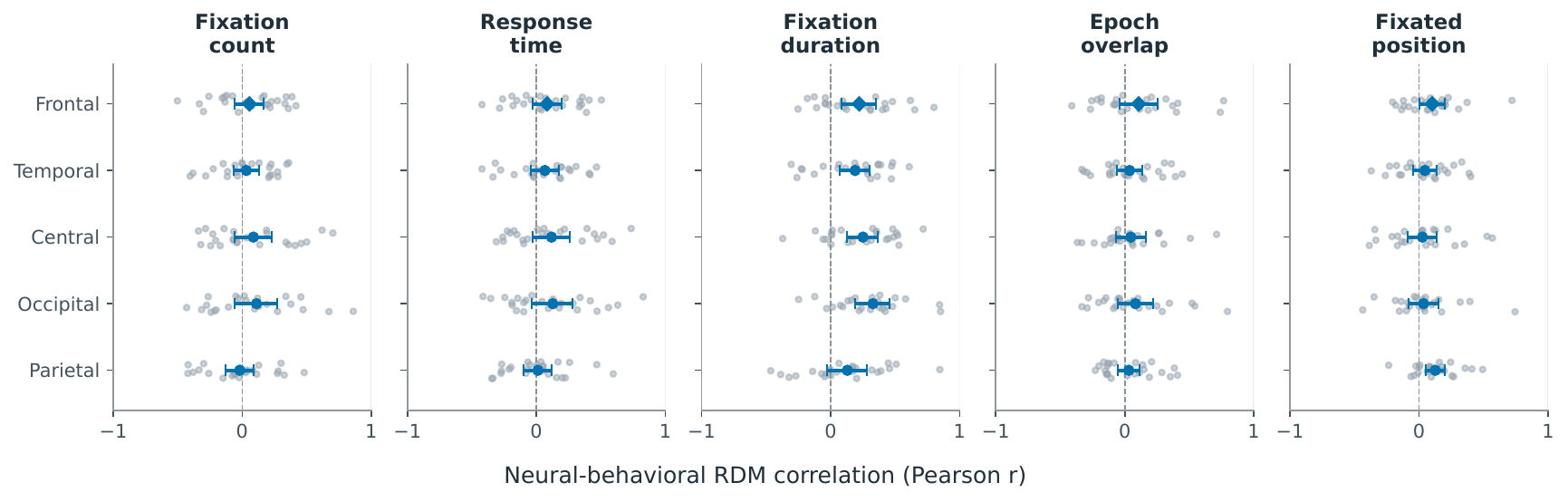}
    \caption{Associations between FRP geometry and fixation or timing measures. Each panel compares participant-level FRP RDMs with RDMs derived from the indicated measure, separately for each regional electrode group. Gray points show individual participants' Pearson correlations ($N=\HierFRPParticipants{}$). Blue markers show correlations averaged in Fisher-$z$ space and back-transformed to $r$; horizontal bars show 95\% Student-$t$ confidence intervals calculated in Fisher-$z$ space and back-transformed to $r$. These intervals are marginal and are not adjusted for multiple comparisons. The associations describe shared pattern-related structure; they do not establish whether these measures explain the FRP--LLM correspondence.}
    \label{fig:nuisance-rdm-results}
\end{figure}

Because FRPs sample EEG during sequence fixations, we next examined whether their representational geometry covaried with gaze behavior and task timing. Fixation statistics can reflect cognitive strategies during reasoning \citep{kucharsky_cognitive_2020,liu_exploring_2025}. We therefore compared FRP RDMs with descriptive RDMs for fixation duration, fixation count, sequence positions, response time, and epoch overlap. Fixation duration showed the strongest association across regions ($r=\HierNuisanceFRPFixationDurationRegionalRangeLow{}$--$\HierNuisanceFRPFixationDurationRegionalRangeHigh{}$), peaking in occipital channels ($r=\HierNuisanceFRPOccipitalFixationDurationRSA{}$) and remaining evident frontally ($r=\HierNuisanceFRPFrontalFixationDurationRSA{}$). Mean fixated sequence position showed a smaller frontal association ($r=\HierNuisanceFRPFrontalPositionRSA{}$), while fixation count, response time, and epoch overlap were generally weaker (\Cref{fig:nuisance-rdm-results}). These associations identify shared structure between fixation statistics and FRPs, but do not establish whether it explains the FRP--LLM correspondence.

\FloatBarrier
\Needspace{8\baselineskip}
\section{Discussion}
By comparing behavior and representational geometry within the same abstract-pattern-completion task, this study examined whether similarities between humans and LLMs could extend beyond task performance. Human participants were more accurate than the LLMs overall, although the largest models approached human accuracy and some models showed relatively human-like pattern-specific difficulty profiles. However, these behavioral measures were not interchangeable: high overall accuracy did not necessarily imply a close match to the human pattern-specific difficulty profile. At the representational level, all LLMs formed representations organized by abstract pattern, with correspondence to the task-design RDM being the strongest in intermediate layers. At these ``task-optimal'' layers, patterns represented similarly by the LLMs also tended to elicit similar frontal FRP responses during human sequence inspection. This correspondence was significant in the pattern-label permutation test, supporting our primary hypothesis of shared representational structure between LLM activations and human frontal FRPs.

Interpreting the LLM–human correspondence requires considering how consistently human representational structure is shared across participants. On average, participants’ RDMs were more similar to group RDMs formed from the other participants for FRPs than for response-locked or pre-trial rest EEG, both across regional electrode groups and within frontal electrodes. This suggests that FRPs captured more consistent representational structure across participants than the two control measures. At the same time, individual-participant LLM RSA was lower than RSA with the participant-averaged group RDM. This difference is expected when group averaging suppresses participant-specific variation, and it means that the group-level correspondence should be interpreted as alignment with a shared component of a heterogeneous human signal rather than as a uniform property of every participant.

Several results further constrain the interpretation of this alignment. Frontal FRP alignment was significant in the exact pattern-label permutation test, supporting correspondence tied to the intended identities of the eight abstract patterns. This makes an arbitrary correspondence between similarly shaped geometries less plausible, but it does not identify the cognitive process responsible for the alignment. The layer-wise profiles place the task-optimal layers in context. Task-design RSA peaked in the middle layers, whereas correspondence with frontal FRPs remained elevated across a broader range extending into later layers. These findings are consistent with task-related representational correspondence, while remaining agnostic about whether the underlying computations are mechanistically shared.

The behavioral and neural results should therefore not be collapsed into a single model ranking. All eight LLMs showed positive FRP RSA when averaged across regions, but their ordering did not simply follow either overall task accuracy or similarity to the human pattern-specific difficulty profile. For example, Qwen2.5-72B and DS-R1-Llama-70B combined high accuracy with relatively human-like pattern-specific difficulty profiles, but they did not have the highest mean FRP RSA across regions. Overall accuracy, similarity in difficulty across patterns, and neural representational correspondence therefore provide related but distinct views of model performance.

The fixation and timing analyses provide complementary information about this task-related activity. Fixation duration showed the strongest association with FRP structure across the regional groups, with correlations ranging from $r=\HierNuisanceFRPFixationDurationRegionalRangeLow{}$ to $r=\HierNuisanceFRPFixationDurationRegionalRangeHigh{}$. The association was strongest in occipital channels and was also evident in frontal channels. Such associations can be cognitively meaningful: fixation behavior has been linked to solution strategies in visual reasoning tasks \citep{kucharsky_cognitive_2020,liu_exploring_2025}.

\subsection{Limitations}
The construction of the FRP measure imposes a temporal limitation. Each trial-level FRP averages several 600-ms EEG epochs following valid fixations on sequence icons. These epochs can overlap in time, so the average does not isolate the EEG response to a single fixation from activity associated with neighboring fixations. FRP, response-locked, and rest EEG were also constructed from different temporal anchors and different numbers of epochs, so their RDMs summarize different portions of the task and should not be treated as identical measurements taken at different moments.

We are relatively certain that the FRP-LLM correspondences across pattern types are not due to remaining electro-ocular artifact still present in the frontal EEG channels because we followed artifact removal procedures \citep{fitzgibbon_automatic_2016} known to remove eye-movements \citep{jung_removing_2000,plochl_combining_2012,fitzgibbon_removal_2007}. Furthermore, occipital and parietal electrode groups, known to reflect visual processing and evidence accumulation \citep{hillyard_eventrelated_1998,oconnell_supramodal_2012} and not electro-ocular artifact, showed similar FRP-LLM correspondences (see \Cref{fig:main-results}D). But it could be that the FRPs are encoding the same cognitive information as the fixation behavior, which is known to be cognitively relevant \citep{kucharsky_cognitive_2020,liu_exploring_2025}, instead of completely \emph{independent} neurocognitive information. Future work should test whether the alignment persists after accounting for fixation behavior.

A more specific interpretive limitation concerns the relationship between abstract-pattern structure and visual complexity. The patterns differ in the number of unique icons and in their repetition and spatial arrangement, so the present analyses do not isolate contributions of these properties to the shared representational geometry. The frontal pairwise contribution plot provides some descriptive context (\suppfigref{fig:matched-rdm-pair-contributions}): positive FRP contributions occur for pairs with the same number of unique icons, including ABABCDCD--ABBACDDC and ABABCDCD--ABCDDCBA, which each contain four. However, these are contributions to the full RDM correlation, not tests of correspondence after controlling for icon count. They therefore do not rule out an account involving visual complexity. A targeted follow-up could compare patterns while holding icon count and other visual properties constant, or explicitly model these properties alongside abstract-pattern structure.

\subsection{Conclusions}
Taken together, the study shows that LLMs can organize sequences of items according to their abstract pattern structure, and that the layers expressing this organization can resemble a reproducible component of human frontal EEG during active sequence inspection. The findings do not establish that LLMs and people use the same reasoning mechanism.

\section{Methods}
\subsection*{Task design}
    In each trial, participants viewed a sequence of icons arranged according to a specific underlying pattern (e.g., “ABBACDDC”). 
    Eight unique patterns were used in the experiment:
    \begin{center}
        \begin{tabular}{ c c c c}
         AAABAAAB & ABBACDDC & ABCDDCBA & ABABCDCD \\ 
         ABBCABBC & ABCDEEDC & ABBAABBA & ABCAABCA
        \end{tabular}
    \end{center}

    \noindent For each trial, the final icon in the sequence was replaced by a question mark. Participants were then required to select the icon from a set of four multiple-choice options that correctly continued the sequence. Participants responded by pressing one of four computer keyboard buttons corresponding to each response. This design was adapted to a text-based version for LLMs, which were presented with sequences of one-word labels describing the icons.
    
    Participants were seated 60 cm from a computer monitor with their head stabilized on an adjustable chin rest to minimize movement, while EEG signals were recorded via a 64-electrode cap and gaze data were captured simultaneously with an eye tracker.
    The icons of a sequence were first individually presented for 600 ms in a randomized order (but at their respective location in the sequence) on the top part of the screen. This ``encoding phase'' was included in the experiment so that participants could register the visual features of every icon without yet attempting to solve the pattern. Subsequently, four choice icons were displayed in a similar manner, but on the lower part of the screen. After these separate presentations, the entire sequence and the four options were presented simultaneously on the screen and remained until the participant responded by pressing a key or until a maximum duration of 12 seconds was reached. We refer to this second phase as the ``decision phase''. By temporally separating these two phases, we aimed to isolate the neural signals related to reasoning processes from those associated with more basic perceptual processing.
    
    The full experiment consisted of 400 unique trials divided into 5 sessions, with 50 trials per pattern type. Each session was divided into 4 blocks of 20 trials, for a total of 80 trials with 10 trials per pattern type, presented in random order.
    
\subsection*{Participants}
Twenty-five healthy adults were recruited from the University's participant pool through an online advertisement. The advertisement specified the eligibility criteria ($\ge$ 18 years; normal or corrected-to-normal vision; no personal or familial history of epilepsy) and the study logistics: five sessions of about one to two hours each. Volunteers were offered either \texteuro100 or six course-credit units as compensation for the completion of the whole study. Five participants withdrew before completing the whole experiment: three withdrew after their first session, one after two sessions, and another after three sessions. Partial datasets were retained for analyses in which the required recordings and pattern coverage were available.

To ensure high-quality EEG and eye tracking recordings, prospective participants were asked to (i) remove mascara, (ii) wear soft contact lenses if possible (when vision correction was needed), (iii) tie long hair in a low ponytail, (iv) arrive with product-free hair. All participants gave written informed consent, and the protocol was approved by the Faculty Ethics Review Board.

\phantomsection\label{sec:sample-inclusion}
\noindent\textit{Sample inclusion.} Participants contributed 108 recorded sessions. Behavioral analyses used the \BehSessions{} sessions from \BehParticipants{} participants represented in both the response-locked and pre-trial rest EEG analysis exports, comprising 8,400 trials. Three recorded sessions were absent from these exports.

Of these \BehSessions{} sessions, 104 were represented in the FRP trial export. Constructing a session-level FRP RDM required at least one trial with a valid FRP feature vector for every abstract pattern in each regional electrode group. Eight sessions failed this criterion, leaving \HierMatchedCompleteSessions{} sessions from \HierMatchedParticipants{} participants. These sessions also had response-locked and rest RDMs and formed the matched sample for both the primary neural comparison and the descriptive layer-wise profiles. The condition-specific reliability and individual-participant analyses used \HierFRPParticipants{} participants and \HierFRPCompleteSessions{} sessions for FRPs, and \HierResponseParticipants{} participants and \HierResponseSessions{} sessions for response-locked and rest EEG. Group RDMs were constructed by averaging sessions equally within participants, then averaging participants equally.

\subsection*{LLMs}
Eight open-weight LLMs were downloaded and run locally using the Hugging Face Transformers library \citep{wolf_transformers_2020}: Phi-4, Gemma-2-2B, Gemma-2-9B, Gemma-2-27B, Llama-3.2-3B, Llama-3.3-70B, Qwen2.5-72B, DS-R1-Llama-70B (see \Cref{tab:LLMs_list} for more details). The models were queried 400 different times, that is, one trial at a time with no context carried over between prompts (i.e., no memory of previous trials), and each of their layers' outputs were extracted.

\begin{table}[ht]
    \resizebox{0.95\textwidth}{!}{%
    \begin{tabular}{llll}
        \toprule
        Full identifier & Simplified identifier & Parameter Count (billions) \\
        \midrule
        \href{https://huggingface.co/microsoft/phi-4}{microsoft/phi-4} & Phi-4 & 14 \\
        \href{https://huggingface.co/google/gemma-2-2b-it}{google/gemma-2-2b-it} & Gemma-2-2B & 2 \\
        \href{https://huggingface.co/google/gemma-2-9b-it}{google/gemma-2-9b-it} & Gemma-2-9B & 9 \\
        \href{https://huggingface.co/google/gemma-2-27b-it}{google/gemma-2-27b-it} & Gemma-2-27B & 27 \\
        \href{https://huggingface.co/meta-llama/Llama-3.2-3B-Instruct}{meta-llama/Llama-3.2-3B-Instruct} & Llama-3.2-3B & 3 \\
        \href{https://huggingface.co/meta-llama/Llama-3.3-70B-Instruct}{meta-llama/Llama-3.3-70B-Instruct} & Llama-3.3-70B & 70 \\
        \href{https://huggingface.co/Qwen/Qwen2.5-72B-Instruct}{Qwen/Qwen2.5-72B-Instruct} & Qwen2.5-72B & 72 \\
        \href{https://huggingface.co/deepseek-ai/DeepSeek-R1-Distill-Llama-70B}{deepseek-ai/DeepSeek-R1-Distill-Llama-70B} & DS-R1-Llama-70B & 70 \\
        \bottomrule
    \end{tabular}
    }
    \caption{Open-weight LLMs evaluated in this study. The ``Full identifier'' column gives the full identifier used on the Hugging Face Hub \citep{huggingface_model_hub_nd}, while the ``Simplified identifier'' column gives the shortened version used for each LLM throughout this paper.}
    \label{tab:LLMs_list}
\end{table}

\subsection*{LLM prompt}
A one-shot prompting strategy was employed, where an unrelated example and its associated answer were given to the LLM before showing it the actual problem to solve. Below is an example of a prompt used to query the LLMs:

\begin{quote}
    \raggedright
    {\fontsize{9pt}{9pt}\selectfont %
    You will be presented with a sequence of words that follow a logical order, but one word in this sequence is missing, indicated by a question mark (?). Your task is to identify the missing word from a given set of options. To successfully complete this task, you should:
    \begin{enumerate}
        \item Analyze the sequence to understand the underlying logical pattern. Pay attention to the order of words, any repetitions, and how each word relates to the others in the sequence.
        \item Do not rely on external tools or databases to analyze the sequence. Your reasoning should be based solely on the internal logic of the sequence as presented.
        \item Consider the provided options carefully. There is only one correct answer that fits logically into the sequence in place of the question mark.
        \item Present your answer in a clear and concise format: 'Answer: [chosen word]' (without the brackets). Include only your final choice in your response, without any additional explanation or text.
    \end{enumerate}
    Your goal is to determine which option logically completes the sequence. Remember, the key to solving this puzzle is understanding the pattern that links the words in the sequence. Use this pattern to decide which of the options fits as the missing word. \par\medskip
    Here is an example: \par
    Sequence: smile eye smile eye smile ? \par
    Options: camera eye bone smile \par
    Answer: eye \par
    
    \noindent Here is the puzzle you must now solve:  \par
    Sequence: star star star guitar star star star ?  \par
    Options: truck star cube guitar}
\end{quote}

\subsection*{EEG apparatus}
EEG data were collected using a 64-electrode headcap from BioSemi \citep{biosemi_headcaps_nd}, arranged according to the international 10-20 system and connected to an EEG amplifier system with a sampling rate of 2048 Hz.
  
\subsection*{Eye tracking apparatus}
Eye movements were recorded from the participants’ dominant eye using an EyeLink 1000 Plus system \citep{srresearch_eyelink1000plus_nd} with a sampling rate of 2000 Hz.

\subsection*{Data analysis}
\subsubsection*{Behavioral analysis}
Behavioral summaries used all task trials from the \BehSessions{} sessions represented in both the response-locked and pre-trial rest EEG exports (\BehParticipants{} participants; see \hyperref[sec:sample-inclusion]{Sample inclusion}). This sample differs from the smaller complete-case sample used for the matched neural comparison. Overall human accuracy was calculated within each participant across their included trials and then averaged equally across participants; the reported standard deviation describes variation across participant accuracies. For the human difficulty profile, accuracy was calculated within each participant and abstract-pattern category, then averaged equally across participants within each of the eight patterns. Pointwise 95\% confidence intervals for this profile were the 2.5th and 97.5th percentiles of 10,000 participant-bootstrap resamples, using the same participant draw across all patterns.

For each LLM, we calculated overall accuracy and accuracy within each pattern. Model--human profile similarity was the Pearson correlation between the model's eight pattern accuracies and the equally weighted human pattern means. Summary means and standard deviations for model accuracy describe the eight fixed models.

\subsubsection*{EEG preprocessing}
EEG preprocessing was performed using the MNE-Python library \citep{gramfort_meg_2013} with the following steps:
\begin{enumerate}
    \item Initial average re-referencing \\
    Noisy or unreliable electrodes were manually flagged during data acquisition or subsequent inspection. The continuous EEG was first re-referenced to the average of the electrodes not marked as bad.
    \item Notch filter (50–250 Hz in 50 Hz steps) \\
    The European electricity grid (Continental Europe Synchronous Area) operates at a standard frequency of 50 Hz. This creates noise at 50 Hz and multiples of it (harmonics) that gets mixed in with the EEG signal. We therefore used a series of narrow, ``notch” filters centred on 50-Hz increments (from 50 to 250 Hz) to target and remove this noise while leaving neighbouring frequencies untouched. Each notch was implemented with MNE’s default zero-phase, overlap-add finite impulse response (FIR) routine. Notch width followed the package default of frequency / 200 (e.g., 0.25 Hz at 50 Hz) with a fixed 1 Hz transition band. Filter length was chosen automatically (6.6 divided by the shortest transition bandwidth, using the default Hamming window), and edges were handled with MNE’s ``reflect\_limited" padding, which mirrors the signal at each edge and then pads with zeros if necessary.
    \item Independent component analysis (ICA) for artifact removal \\
    To isolate non-neural sources such as eye blinks, saccades, or muscle movements, we duplicated the notch-filtered data, applied a broad 1–100 Hz band-pass filter to this copy, and fitted an extended Infomax ICA \citep{langlois_introduction_2010}. The resulting components were automatically labelled with the MNE-ICALabel classifier \citep{li_mne-icalabel_2022}. Only components labelled ``brain” were retained; all other components were excluded. The fitted ICA solution was then applied to the notch-filtered continuous data.
    \item Band-pass filter (0.1–100 Hz) \\
    After ICA, a final 0.1–100 Hz band-pass filter was applied to the continuous data.
    \item Interpolation of manually identified bad electrodes \\
    Electrodes marked as bad were then interpolated using MNE's spherical-spline interpolation \citep{perrin_spherical_1989}.
    \item Final average re-referencing \\
    Following interpolation, the data were re-referenced once more to the average across electrodes.
\end{enumerate}

\subsubsection*{LLMs' layer activations on sequence tokens}

LLMs process text in the form of discrete tokens: small chunks of text such as whole words or, more often, sub-word pieces (e.g., “guit”, “ar”). Each token is replaced by an integer ID drawn from the model’s vocabulary, which is then turned into a continuous embedding vector (a fixed-length list of real numbers that encodes the token’s learned semantic and contextual properties). Most modern LLMs pass these embeddings up through a stack of identical transformer blocks, also referred to as hidden layers. Each layer outputs a hidden-state matrix whose rows correspond to the tokens and whose columns hold feature values computed at that layer. We extracted a subset of these values from the output of every layer to isolate those specifically associated with the tokens of a given abstract sequence. For example, using the following prompt:
\begin{quote}
    Here is the puzzle you must now solve:\\
    Sequence: star star star guitar star star star ?\\
    Options: truck star cube guitar
\end{quote}
only the activations for the tokens that make up the ``Sequence" line: (i.e., `star', `star', `star', `guitar', `star', `star', `star', `?') are extracted, thus producing a trial-level representation of a layer's activity related to the abstract sequence.

\subsubsection*{Representational similarity analysis (RSA)}
\label{sec:rsa}

\textit{Analytic question and hierarchy.} The central question was whether the representational relationships among the eight abstract patterns in LLM layers resembled the corresponding relationships in human EEG. The primary analysis was deliberately hierarchical: trials, sessions, participants, and the group remained distinct levels rather than being pooled immediately.
The steps were as follows.

\begin{enumerate}
    \item \textit{Inputs and regional views.} FRPs, response-locked EEG, and pre-trial rest EEG were analysed across five non-overlapping regional channel groups (\suppfigref{fig:eeg-channel-groups}): frontal (17 electrodes), central (14), temporal (6), parietal (14), and occipital (13), using standardized 10--10 nomenclature and spatial adjacency. These are non-overlapping groups that together cover the 64-electrode montage.
    \item \textit{Trial-level EEG features.} FRPs were derived from eye-tracker-labelled fixations during the decision phase, when the full sequence and response options were visible. Fixations whose mean gaze position lay within one of the eight sequence-icon bounding boxes were retained; fixations on response options were excluded. For each retained fixation, 600 ms of continuously preprocessed EEG beginning at fixation onset was extracted, without fixation-specific baseline correction. Retained sequence-fixation epochs were averaged within a trial, yielding one trial-level FRP that reflects ongoing self-paced sequence inspection rather than an isolated response to one fixation. Response-locked EEG corresponds to the 600-ms window preceding the button press (or the end of the response window for omissions), whereas pre-trial rest EEG corresponds to the 600 ms preceding the trial-start marker, during fixation-cross presentation. For every EEG type and regional group, electrodes were averaged at each time point, yielding a trial-level time-course feature vector. Descriptive grand-average channel waveforms and frontal pattern-mean FRPs for the matched sample are shown in \suppfigref{fig:matched-frp-waveforms}.
    \item \textit{Session RDMs.} Within each session and abstract-pattern class, trial feature vectors were averaged, producing eight session-specific pattern vectors. A session-specific $8\times8$ correlation-distance RDM ($1-r$) was then calculated from these vectors.
    \item \textit{Participant and group RDMs.} Each participant's session RDMs were averaged, and the resulting participant RDMs were then averaged to form a \emph{participant-averaged group RDM}. 
    \item \label{meth:llm-rdms}\textit{LLM RDMs.} Correlation-distance RDMs were calculated from sequence-token hidden-layer activations for every LLM layer. Task-optimal layers were selected by maximizing the Pearson correlation between each layer's trial-level RDM and a task-design RDM that encoded within-pattern similarity and between-pattern dissimilarity, independently of EEG data. For each model, activations in the selected layer were then averaged within the eight pattern classes to form one eight-pattern LLM RDM. Descriptive layer-wise plots repeated this comparison over normalized layer depth. Layers are numbered from one in the manuscript and figures; normalized depth is the layer number divided by the total number of transformer layers. Stored analysis indices remain zero-based. All RDM construction and comparisons used rsatoolbox \citep{rsatoolbox_documentation_nd}, with correlation distance and Pearson RDM similarity.
    \item \textit{Primary LLM--human RSA and reporting.} Each LLM RDM was correlated separately with each participant-averaged group RDM using the 28 unique upper-triangular RDM entries. The eight model-specific values and, for compact descriptive summaries, their mean are reported for each EEG type and regional view. The mean across the eight preselected models describes this fixed model set; it is not treated as a random sample from an unspecified population of LLMs.
    \item \label{meth:individual-reliability}\textit{Individual-participant and reliability analyses.} As supporting analyses, each participant RDM was compared with each fixed LLM RDM and the resulting values were summarized across participants. Separately, each participant RDM was correlated with a group RDM constructed from all \emph{other} participants (leave-one-participant-out; LOO).
    \item \textit{Fixation and timing correlates.} For these descriptive analyses, each trial-level variable was averaged within session and pattern class. The absolute difference between every pair of the eight pattern means formed a session RDM for that measure, and these RDMs were averaged within participant. Each participant's RDM for each measure was correlated with the corresponding participant-level neural RDM. The \HierNuisanceDiagnosticCount{} measures were valid fixation count, response time, mean fixation duration, adjacent-epoch overlap, and mean fixated stimulus position. Adjacent-epoch overlap was the proportion of consecutive fixation pairs whose onset-to-onset interval was shorter than 600 ms, meaning that their 600-ms EEG epochs overlapped. Mean fixated stimulus position was the average sequence-position index of the fixated icons (the stimulus position within the displayed sequence), not a raw screen-coordinate measure. Values came from FRP trial metadata and were joined to response and rest trials by participant, session, and trial number, so the measure definitions were held constant across EEG measures. These descriptive correlations quantify shared pattern-related structure; they do not identify its cognitive source or establish causal relationships. Retained participants, sessions, trials, and valid fixation counts were summarized by EEG measure and pattern.
    \item \label{meth:pattern-label-test}\textit{Exact pattern-label permutation test.} To test the primary hypothesis of pattern-specific alignment between LLM representations and frontal FRPs, LLM RDMs were held fixed while the rows and matching columns of the participant-averaged group RDM were jointly permuted. All \HierPatternLabelPermutationCount{} possible mappings of the eight pattern labels were enumerated. The same mapping was applied to every regional and EEG RDM within a permutation, preserving their shared dependence. The statistic was
    \[
        T=\frac{1}{\HierLLMCount{}G}\sum_{m=1}^{\HierLLMCount{}}\sum_{g=1}^{G}\operatorname{arctanh}(r_{mg}),
    \]
    where $r_{mg}$ is the Pearson RDM correlation for model $m$ and region $g$, and $G=\HierChannelGroupCount{}$ for the overall test or $G=1$ for the frontal test. Exact one-sided p-values were the proportion of all \HierPatternLabelPermutationCount{} label permutations, including the original mapping, whose statistic was at least as large as the observed statistic. Thus, inference used the permutation distribution of $T$, rather than a parametric $t$ or standard-normal test. Frontal FRP alignment was the single primary neural hypothesis, for which we report the unadjusted exact one-sided permutation p-value. Frontal response-locked and pre-trial rest EEG were evaluated as secondary controls, with their unadjusted permutation p-values reported separately. As a supplementary robustness check, we also report Holm-adjusted p-values \citep{holm_sequential_1979} treating all three frontal EEG-condition tests as one family; this broader correction does not define the primary-hypothesis decision. The three exploratory across-region EEG-condition tests were treated as a separate family and corrected using Holm's procedure. This null distribution tests whether the observed LLM--human match is greater than values obtained after breaking the intended pattern-label mapping.
\end{enumerate}

\section{Data and code availability}

All data and code required to reproduce the study's findings are openly available. The anonymized behavioral, raw EEG, and eye-tracking datasets, as well as the LLM responses and activations, are available on Figshare (\url{https://doi.org/10.21942/uva.29573534}). The raw EEG data follow the Brain Imaging Data Structure (BIDS), using its EEG extension (EEG-BIDS) \citep{pernet_eeg-bids_2019}. Preprocessed EEG files are available upon request. The analysis pipeline, figure generation scripts, and stimuli are available on GitHub (\url{https://github.com/chris-pinier/abstract_reasoning}).

\section{Funding}
The Dutch government and the University of Amsterdam provided a starting grant supporting joint work by Christopher Pinier, Claire E. Stevenson, and Michael D. Nunez. This research was also funded in part by the Dutch Research Council (NWO) Talent Programme Vidi project ``How smart is AI really? Methods to investigate intelligence in humans and AI'' awarded to Claire E. Stevenson with file number VI.Vidi.241G.063 under the grant \url{https://doi.org/10.61686/VCCQU27680}.

\begingroup
\NewCommandCopy{\paperBibliographyHref}{\href}
\ExplSyntaxOn
\RenewDocumentCommand{\href}{ O{} m m }
  {
    \regex_match:nnTF { \A 10 \. [0-9]{4,9} / } {#2}
      { \paperBibliographyHref[#1]{https\c_colon_str//doi.org/#2}{#3} }
      { \paperBibliographyHref[#1]{#2}{#3} }
  }
\ExplSyntaxOff
\bibliography{references}

\begin{thebibliography}{64}
\providecommand{\natexlab}[1]{#1}
\providecommand{\urlprefix}{}
\providecommand{\doiprefix}{doi: }

\bibitem[{Badre et~al.(2010)Badre, David and Kayser, Andrew S. and D'Esposito,
  Mark}]{badre_frontal_2010}
\textbf{\color{eLifeMediumGrey} Badre D}, Kayser AS, D'Esposito M.
\newblock Frontal cortex and the discovery of abstract action rules.
\newblock Neuron.  2010 Apr; 66(2):315--326.
\newblock
  \urlprefix\url{https://www.cell.com/neuron/abstract/S0896-6273(10)00196-0},
  \href{10.1016/j.neuron.2010.03.025}{\doiprefix
  \detokenize{10.1016/j.neuron.2010.03.025}}, publisher: Elsevier.

\bibitem[{Balunovi{\'c} et~al.(2025)Balunovi{\'c}, Mislav and Dekoninck, Jasper
  and Petrov, Ivo and Jovanovi{\'c}, Nikola and Vechev,
  Martin}]{balunovic_matharena_2025}
\textbf{\color{eLifeMediumGrey} Balunovi{\'c} M}, Dekoninck J, Petrov I,
  Jovanovi{\'c} N, Vechev M.
\newblock {MathArena}: Evaluating {LLMs} on Uncontaminated Math Competitions.
\newblock In: \emph{Advances in Neural Information Processing Systems},
  vol.~38; 2025. \href{10.52202/085713-0679}{\doiprefix
  \detokenize{10.52202/085713-0679}}.

\bibitem[{{BioSemi}(n.d.)}]{biosemi_headcaps_nd}
\textbf{\color{eLifeMediumGrey} {BioSemi}}, Headcap, headcaps; n.d.
\newblock \urlprefix\url{https://www.biosemi.com/headcap.htm}, accessed 1
  October 2026.

\bibitem[{Bowers et~al.(2023)Bowers, Jeffrey S. and Malhotra, Gaurav and
  Dujmović, Marin and Llera Montero, Milton and Tsvetkov, Christian and
  Biscione, Valerio and Puebla, Guillermo and Adolfi, Federico and Hummel, John
  E. and Heaton, Rachel F. and Evans, Benjamin D. and Mitchell, Jeffrey and
  Blything, Ryan}]{bowers_deep_2022}
\textbf{\color{eLifeMediumGrey} Bowers JS}, Malhotra G, Dujmović M,
  Llera~Montero M, Tsvetkov C, Biscione V, Puebla G, Adolfi F, Hummel JE,
  Heaton RF, Evans BD, Mitchell J, Blything R.
\newblock Deep {Problems} with {Neural} {Network} {Models} of {Human} {Vision}.
\newblock Behavioral and Brain Sciences.  2023 Dec; 46:e385.
\newblock \href{10.1017/S0140525X22002813}{\doiprefix
  \detokenize{10.1017/S0140525X22002813}}.

\bibitem[{Bubeck et~al.(2023)Bubeck, Sébastien and Chandrasekaran, Varun and
  Eldan, Ronen and Gehrke, Johannes and Horvitz, Eric and Kamar, Ece and Lee,
  Peter and Lee, Yin Tat and Li, Yuanzhi and Lundberg, Scott and Nori, Harsha
  and Palangi, Hamid and Ribeiro, Marco Tulio and Zhang,
  Yi}]{bubeck_sparks_2023}
\textbf{\color{eLifeMediumGrey} Bubeck S}, Chandrasekaran V, Eldan R, Gehrke J,
  Horvitz E, Kamar E, Lee P, Lee YT, Li Y, Lundberg S, Nori H, Palangi H,
  Ribeiro MT, Zhang Y.
\newblock Sparks of artificial general intelligence: early experiments with
  {GPT}-4; 2023, \urlprefix\url{http://arxiv.org/abs/2303.12712},
  \href{10.48550/arXiv.2303.12712}{\doiprefix
  \detokenize{10.48550/arXiv.2303.12712}}, arXiv:2303.12712 [cs].

\bibitem[{Caucheteux and King(2022)Caucheteux, Charlotte and King,
  Jean-Rémi}]{caucheteux_brains_2022}
\textbf{\color{eLifeMediumGrey} Caucheteux C}, King JR.
\newblock Brains and algorithms partially converge in natural language
  processing.
\newblock Communications Biology.  2022 Feb; 5(1):134.
\newblock \urlprefix\url{https://www.nature.com/articles/s42003-022-03036-1},
  \href{10.1038/s42003-022-03036-1}{\doiprefix
  \detokenize{10.1038/s42003-022-03036-1}}, publisher: Nature Publishing Group.

\bibitem[{Caudle et~al.(2023)Caudle, M. M. and Spadoni, A. D. and Schiehser, D.
  M. and Simmons, A. N. and Bomyea, J.}]{caudle_neural_2023}
\textbf{\color{eLifeMediumGrey} Caudle MM}, Spadoni AD, Schiehser DM, Simmons
  AN, Bomyea J.
\newblock Neural activity and network analysis for understanding reasoning
  using the matrix reasoning task.
\newblock Cognitive Processing.  2023; 24(4):585--594.
\newblock
  \urlprefix\url{https://www.ncbi.nlm.nih.gov/pmc/articles/PMC10533635/},
  \href{10.1007/s10339-023-01152-2}{\doiprefix
  \detokenize{10.1007/s10339-023-01152-2}}.

\bibitem[{Chen et~al.(2026)Chen, Kedi and Ruan, Dezhao and Dan, Yuhao and Wang,
  Yaoting and Yan, Siyu and Wu, Xuecheng and Zhang, Yinqi and Chen, Qin and
  Zhou, Jie and He, Liang and Qi, Biqing and Li, Linyang and Guo, Qipeng and
  Shi, Xiaoming and Zhang, Wei}]{chen_survey_2026}
\textbf{\color{eLifeMediumGrey} Chen K}, Ruan D, Dan Y, Wang Y, Yan S, Wu X,
  Zhang Y, Chen Q, Zhou J, He L, Qi B, Li L, Guo Q, Shi X, Zhang W.
\newblock A Survey of Inductive Reasoning for Large Language Models.
\newblock In: Liakata M, Moreira VP, Zhang J, Jurgens D, editors.
  \emph{Proceedings of the 64th {{Annual Meeting}} of the {{Association}} for
  {{Computational Linguistics}} ({{Volume}} 1: {{Long Papers}})} San Diego,
  California, United States: Association for Computational Linguistics; 2026.
  p. 31356--31382.
\newblock \urlprefix\url{https://aclanthology.org/2026.acl-long.1447/},
  \href{10.18653/v1/2026.acl-long.1447}{\doiprefix
  \detokenize{10.18653/v1/2026.acl-long.1447}}.

\bibitem[{Choi et~al.(2008)Choi, Yu Yong and Shamosh, Noah A. and Cho, Sun Hee
  and DeYoung, Colin G. and Lee, Min Joo and Lee, Jong-Min and Kim, Sun I. and
  Cho, Zang-Hee and Kim, Kyungjin and Gray, Jeremy R. and Lee, Kun
  Ho}]{choi_multiple_2008}
\textbf{\color{eLifeMediumGrey} Choi YY}, Shamosh NA, Cho SH, DeYoung CG, Lee
  MJ, Lee JM, Kim SI, Cho ZH, Kim K, Gray JR, Lee KH.
\newblock Multiple bases of human intelligence revealed by cortical thickness
  and neural activation.
\newblock Journal of Neuroscience.  2008 Oct; 28(41):10323--10329.
\newblock
  \urlprefix\url{https://www.ncbi.nlm.nih.gov/pmc/articles/PMC6671030/},
  \href{10.1523/JNEUROSCI.3259-08.2008}{\doiprefix
  \detokenize{10.1523/JNEUROSCI.3259-08.2008}}.

\bibitem[{Chuderski(2022)Chuderski, Adam}]{chuderski_fluid_2022}
\textbf{\color{eLifeMediumGrey} Chuderski A}.
\newblock Fluid intelligence emerges from representing relations.
\newblock Journal of Intelligence.  2022 Aug; 10(3):51.
\newblock
  \urlprefix\url{https://www.ncbi.nlm.nih.gov/pmc/articles/PMC9396997/},
  \href{10.3390/jintelligence10030051}{\doiprefix
  \detokenize{10.3390/jintelligence10030051}}.

\bibitem[{Davis et~al.(2017)Davis, Tyler and Goldwater, Micah and Giron,
  Josue}]{davis_concrete_2017}
\textbf{\color{eLifeMediumGrey} Davis T}, Goldwater M, Giron J.
\newblock From concrete examples to abstract relations: the rostrolateral
  prefrontal cortex integrates novel examples into relational categories.
\newblock Cerebral Cortex.  2017 Apr; 27(4):2652--2670.
\newblock \urlprefix\url{https://doi.org/10.1093/cercor/bhw099},
  \href{10.1093/cercor/bhw099}{\doiprefix \detokenize{10.1093/cercor/bhw099}}.

\bibitem[{Degno and Liversedge(2020)Degno, Federica and Liversedge, Simon
  P.}]{degno_eye_2020}
\textbf{\color{eLifeMediumGrey} Degno F}, Liversedge SP.
\newblock Eye movements and fixation-related potentials in reading: a review.
\newblock Vision.  2020 Feb; 4(1):11.
\newblock
  \urlprefix\url{https://www.ncbi.nlm.nih.gov/pmc/articles/PMC7157570/},
  \href{10.3390/vision4010011}{\doiprefix \detokenize{10.3390/vision4010011}}.

\bibitem[{Dima et~al.(2024)Dima, Diana C. and Janarthanan, Sugitha and Culham,
  Jody C. and Mohsenzadeh, Yalda}]{dima_shared_2024}
\textbf{\color{eLifeMediumGrey} Dima DC}, Janarthanan S, Culham JC, Mohsenzadeh
  Y.
\newblock Shared representations of human actions across vision and language.
\newblock Neuropsychologia.  2024 Sep; 202:108962.
\newblock
  \urlprefix\url{https://www.sciencedirect.com/science/article/pii/S0028393224001775},
  \href{10.1016/j.neuropsychologia.2024.108962}{\doiprefix
  \detokenize{10.1016/j.neuropsychologia.2024.108962}}.

\bibitem[{Doerig et~al.(2024)Doerig, Adrien and Kietzmann, Tim C. and Allen,
  Emily and Wu, Yihan and Naselaris, Thomas and Kay, Kendrick and Charest,
  Ian}]{doerig_visual_2024}
\textbf{\color{eLifeMediumGrey} Doerig A}, Kietzmann TC, Allen E, Wu Y,
  Naselaris T, Kay K, Charest I.
\newblock Visual representations in the human brain are aligned with large
  language models; 2024, \urlprefix\url{http://arxiv.org/abs/2209.11737},
  \href{10.48550/arXiv.2209.11737}{\doiprefix
  \detokenize{10.48550/arXiv.2209.11737}}, arXiv:2209.11737 version: 2.

\bibitem[{Dumontheil(2014)Dumontheil, Iroise}]{dumontheil_development_2014}
\textbf{\color{eLifeMediumGrey} Dumontheil I}.
\newblock Development of abstract thinking during childhood and adolescence:
  the role of rostrolateral prefrontal cortex.
\newblock Developmental Cognitive Neuroscience.  2014 Oct; 10:57--76.
\newblock
  \urlprefix\url{https://www.sciencedirect.com/science/article/pii/S1878929314000516},
  \href{10.1016/j.dcn.2014.07.009}{\doiprefix
  \detokenize{10.1016/j.dcn.2014.07.009}}.

\bibitem[{Duncan(2010)Duncan, John}]{duncan_multiple-demand_2010}
\textbf{\color{eLifeMediumGrey} Duncan J}.
\newblock The multiple-demand ({MD}) system of the primate brain: mental
  programs for intelligent behaviour.
\newblock Trends in Cognitive Sciences.  2010 Apr; 14(4):172--179.
\newblock \href{10.1016/j.tics.2010.01.004}{\doiprefix
  \detokenize{10.1016/j.tics.2010.01.004}}.

\bibitem[{Feghhi et~al.(2024)Feghhi, Ebrahim and Hadidi, Nima and Song, Bryan
  and Blank, Idan A. and Kao, Jonathan C.}]{feghhi_what_2024}
\textbf{\color{eLifeMediumGrey} Feghhi E}, Hadidi N, Song B, Blank IA, Kao JC.
\newblock What {Are} {Large} {Language} {Models} {Mapping} to in the {Brain}?
  {A} {Case} {Against} {Over}-{Reliance} on {Brain} {Scores}; 2024,
  \urlprefix\url{http://arxiv.org/abs/2406.01538},
  \href{10.48550/arXiv.2406.01538}{\doiprefix
  \detokenize{10.48550/arXiv.2406.01538}}, arXiv:2406.01538 [cs] version: 1.

\bibitem[{Ferrer et~al.(2009)Ferrer, Emilio and O'Hare, Elizabeth D. and Bunge,
  Silvia A.}]{ferrer_fluid_2009}
\textbf{\color{eLifeMediumGrey} Ferrer E}, O'Hare ED, Bunge SA.
\newblock Fluid reasoning and the developing brain.
\newblock Frontiers in Neuroscience.  2009 May; 3(1):46--51.
\newblock \href{10.3389/neuro.01.003.2009}{\doiprefix
  \detokenize{10.3389/neuro.01.003.2009}}.

\bibitem[{Fitzgibbon et~al.(2016)Fitzgibbon, S. P. and DeLosAngeles, D. and
  Lewis, T. W. and Powers, D. M. W. and Grummett, T. S. and Whitham, E. M. and
  Ward, L. M. and Willoughby, J. O. and Pope, K.
  J.}]{fitzgibbon_automatic_2016}
\textbf{\color{eLifeMediumGrey} Fitzgibbon SP}, DeLosAngeles D, Lewis TW,
  Powers DMW, Grummett TS, Whitham EM, Ward LM, Willoughby JO, Pope KJ.
\newblock Automatic determination of {EMG}-contaminated components and
  validation of independent component analysis using {EEG} during pharmacologic
  paralysis.
\newblock Clinical Neurophysiology.  2016; 127(3):1781--1793.
\newblock \href{10.1016/j.clinph.2015.12.009}{\doiprefix
  \detokenize{10.1016/j.clinph.2015.12.009}}.

\bibitem[{Fitzgibbon et~al.(2007)Fitzgibbon, S. P. and Powers, D. M. W. and
  Pope, K. J. and Clark, C. R.}]{fitzgibbon_removal_2007}
\textbf{\color{eLifeMediumGrey} Fitzgibbon SP}, Powers DMW, Pope KJ, Clark CR.
\newblock Removal of {EEG} Noise and Artifact Using Blind Source Separation.
\newblock Journal of Clinical Neurophysiology.  2007; 24(3):232--243.
\newblock \href{10.1097/WNP.0b013e3180556926}{\doiprefix
  \detokenize{10.1097/WNP.0b013e3180556926}}.

\bibitem[{Goldstein et~al.(2024)Goldstein, Ariel and Grinstein-Dabush, Avigail
  and Schain, Mariano and Wang, Haocheng and Hong, Zhuoqiao and Aubrey, Bobbi
  and Nastase, Samuel A. and Zada, Zaid and Ham, Eric and Feder, Amir and
  Gazula, Harshvardhan and Buchnik, Eliav and Doyle, Werner and Devore, Sasha
  and Dugan, Patricia and Reichart, Roi and Friedman, Daniel and Brenner,
  Michael and Hassidim, Avinatan and Devinsky, Orrin and Flinker, Adeen and
  Hasson, Uri}]{goldstein_alignment_2024}
\textbf{\color{eLifeMediumGrey} Goldstein A}, Grinstein-Dabush A, Schain M,
  Wang H, Hong Z, Aubrey B, Nastase SA, Zada Z, Ham E, Feder A, Gazula H,
  Buchnik E, Doyle W, Devore S, Dugan P, Reichart R, Friedman D, Brenner M,
  Hassidim A, Devinsky O, et~al.
\newblock Alignment of brain embeddings and artificial contextual embeddings in
  natural language points to common geometric patterns.
\newblock Nature Communications.  2024 Mar; 15(1):2768.
\newblock \urlprefix\url{https://www.nature.com/articles/s41467-024-46631-y},
  \href{10.1038/s41467-024-46631-y}{\doiprefix
  \detokenize{10.1038/s41467-024-46631-y}}, publisher: Nature Publishing Group.

\bibitem[{Gramfort et~al.(2013)Gramfort, Alexandre and Luessi, Martin and
  Larson, Eric and Engemann, Denis A. and Strohmeier, Daniel and Brodbeck,
  Christian and Goj, Roman and Jas, Mainak and Brooks, Teon and Parkkonen,
  Lauri and Hämäläinen, Matti}]{gramfort_meg_2013}
\textbf{\color{eLifeMediumGrey} Gramfort A}, Luessi M, Larson E, Engemann DA,
  Strohmeier D, Brodbeck C, Goj R, Jas M, Brooks T, Parkkonen L, Hämäläinen
  M.
\newblock {MEG} and {EEG} data analysis with {MNE}-python.
\newblock Frontiers in Neuroscience.  2013 Dec; 7:267.
\newblock \href{10.3389/fnins.2013.00267}{\doiprefix
  \detokenize{10.3389/fnins.2013.00267}}.

\bibitem[{Gray et~al.(2003)Gray, Jeremy R. and Chabris, Christopher F. and
  Braver, Todd S.}]{gray_neural_2003}
\textbf{\color{eLifeMediumGrey} Gray JR}, Chabris CF, Braver TS.
\newblock Neural mechanisms of general fluid intelligence.
\newblock Nature Neuroscience.  2003 Mar; 6(3):316--322.
\newblock \urlprefix\url{https://www.nature.com/articles/nn1014},
  \href{10.1038/nn1014}{\doiprefix \detokenize{10.1038/nn1014}}, publisher:
  Nature Publishing Group.

\bibitem[{Hillyard and Anllo-Vento(1998)Hillyard, Steven A. and Anllo-Vento,
  Lourdes}]{hillyard_eventrelated_1998}
\textbf{\color{eLifeMediumGrey} Hillyard SA}, Anllo-Vento L.
\newblock Event-related brain potentials in the study of visual selective
  attention.
\newblock Proceedings of the National Academy of Sciences.  1998;
  95(3):781--787.
\newblock \href{10.1073/pnas.95.3.781}{\doiprefix
  \detokenize{10.1073/pnas.95.3.781}}.

\bibitem[{Holm(1979)Holm, Sture}]{holm_sequential_1979}
\textbf{\color{eLifeMediumGrey} Holm S}.
\newblock A Simple Sequentially Rejective Multiple Test Procedure.
\newblock Scandinavian Journal of Statistics.  1979; 6(2):65--70.
\newblock \urlprefix\url{https://www.jstor.org/stable/4615733}.

\bibitem[{{Hugging Face}(n.d.)}]{huggingface_model_hub_nd}
\textbf{\color{eLifeMediumGrey} {Hugging Face}}, The Model Hub; n.d.
\newblock \urlprefix\url{https://huggingface.co/docs/hub/models-the-hub},
  accessed 1 October 2026.

\bibitem[{Iaia et~al.(2025)Iaia, Cosimo and Choksi, Bhavin and Wiebers, Emily
  and Roig, Gemma and Fiebach, Christian J.}]{iaia_representational_2025}
\textbf{\color{eLifeMediumGrey} Iaia C}, Choksi B, Wiebers E, Roig G, Fiebach
  CJ.
\newblock The representational alignment between humans and language models is
  implicitly driven by a concreteness effect; 2025,
  \urlprefix\url{http://arxiv.org/abs/2505.15682},
  \href{10.48550/arXiv.2505.15682}{\doiprefix
  \detokenize{10.48550/arXiv.2505.15682}}, arXiv:2505.15682 [cs] version: 1.

\bibitem[{Jung et~al.(2000)Jung, Tzyy-Ping and Makeig, Scott and Humphries,
  Colin and Lee, Te-Won and McKeown, Martin J. and Iragui, Vicente and
  Sejnowski, Terrence J.}]{jung_removing_2000}
\textbf{\color{eLifeMediumGrey} Jung TP}, Makeig S, Humphries C, Lee TW,
  McKeown MJ, Iragui V, Sejnowski TJ.
\newblock Removing electroencephalographic artifacts by blind source
  separation.
\newblock Psychophysiology.  2000; 37(2):163--178.
\newblock \href{10.1111/1469-8986.3720163}{\doiprefix
  \detokenize{10.1111/1469-8986.3720163}}.

\bibitem[{Krizhevsky et~al.(2012)Krizhevsky, Alex and Sutskever, Ilya and
  Hinton, Geoffrey E}]{krizhevsky_imagenet_2012}
\textbf{\color{eLifeMediumGrey} Krizhevsky A}, Sutskever I, Hinton GE.
\newblock {ImageNet} {Classification} with {Deep} {Convolutional} {Neural}
  {Networks}.
\newblock In: \emph{Advances in {Neural} {Information} {Processing} {Systems}},
  vol.~25 Curran Associates, Inc.; 2012. p. 1097--1105.
\newblock
  \urlprefix\url{https://proceedings.neurips.cc/paper/2012/hash/c399862d3b9d6b76c8436e924a68c45b-Abstract.html}.

\bibitem[{Kucharský et~al.(2020)Kucharský, {\v{S}}imon and Visser, Ingmar and
  Truțescu, Gabriela-Olivia and Laurence, Paulo G. and Zaharieva, Martina and
  Raijmakers, Maartje E. J.}]{kucharsky_cognitive_2020}
\textbf{\color{eLifeMediumGrey} Kucharský {\v{S}}}, Visser I, Truțescu GO,
  Laurence PG, Zaharieva M, Raijmakers MEJ.
\newblock Cognitive {Strategies} {Revealed} by {Clustering} {Eye} {Movement}
  {Transitions}.
\newblock Journal of Eye Movement Research.  2020 Feb; 13(1):1--20.
\newblock \urlprefix\url{https://www.mdpi.com/1995-8692/13/1/1},
  \href{10.16910/jemr.13.1.1}{\doiprefix \detokenize{10.16910/jemr.13.1.1}}.

\bibitem[{Langlois et~al.(2010)Langlois, Dominic and Chartier, Sylvain and
  Gosselin, Dominique}]{langlois_introduction_2010}
\textbf{\color{eLifeMediumGrey} Langlois D}, Chartier S, Gosselin D.
\newblock An introduction to independent component analysis: {InfoMax} and
  {FastICA} algorithms.
\newblock Tutorials in Quantitative Methods for Psychology.  2010 Mar;
  6(1):31--38.
\newblock \urlprefix\url{http://www.tqmp.org/RegularArticles/vol06-1/p031},
  \href{10.20982/tqmp.06.1.p031}{\doiprefix
  \detokenize{10.20982/tqmp.06.1.p031}}.

\bibitem[{LeCun et~al.(2015)LeCun, Yann and Bengio, Yoshua and Hinton,
  Geoffrey}]{lecun_deep_2015}
\textbf{\color{eLifeMediumGrey} LeCun Y}, Bengio Y, Hinton G.
\newblock Deep learning.
\newblock Nature.  2015 May; 521(7553):436--444.
\newblock \urlprefix\url{https://www.nature.com/articles/nature14539},
  \href{10.1038/nature14539}{\doiprefix \detokenize{10.1038/nature14539}},
  publisher: Nature Publishing Group.

\bibitem[{Lei et~al.(2025)Lei, Yu and Ge, Xingyang and Zhang, Yi and Yang,
  Yiming and Ma, Bolei}]{lei_large_2025}
\textbf{\color{eLifeMediumGrey} Lei Y}, Ge X, Zhang Y, Yang Y, Ma B.
\newblock Do large language models think like the brain? {Sentence}-level
  evidence from {fMRI} and hierarchical embeddings; 2025,
  \urlprefix\url{http://arxiv.org/abs/2505.22563},
  \href{10.48550/arXiv.2505.22563}{\doiprefix
  \detokenize{10.48550/arXiv.2505.22563}}, arXiv:2505.22563 [cs].

\bibitem[{Lewis and Mitchell(2025)Lewis, Martha and Mitchell,
  Melanie}]{lewis_evaluating_2024}
\textbf{\color{eLifeMediumGrey} Lewis M}, Mitchell M.
\newblock Evaluating the Robustness of Analogical Reasoning in {GPT} Models.
\newblock Transactions on Machine Learning Research.  2025;
  \urlprefix\url{https://openreview.net/forum?id=t5cy5v9wph}.

\bibitem[{Li et~al.(2022)Li, Adam and Feitelberg, Jacob and Saini, Anand
  Prakash and Höchenberger, Richard and Scheltienne,
  Mathieu}]{li_mne-icalabel_2022}
\textbf{\color{eLifeMediumGrey} Li A}, Feitelberg J, Saini AP, Höchenberger R,
  Scheltienne M.
\newblock {MNE}-{ICALabel}: automatically annotating {ICA} components with
  {ICLabel} in python.
\newblock Journal of Open Source Software.  2022 Aug; 7(76):4484.
\newblock \urlprefix\url{https://joss.theoj.org/papers/10.21105/joss.04484},
  \href{10.21105/joss.04484}{\doiprefix \detokenize{10.21105/joss.04484}}.

\bibitem[{Liu et~al.(2025)Liu, Yaohui and He, Keren and Man, Kaiwen and Zhan,
  Peida}]{liu_exploring_2025}
\textbf{\color{eLifeMediumGrey} Liu Y}, He K, Man K, Zhan P.
\newblock Exploring {Critical} {Eye}-{Tracking} {Metrics} for {Identifying}
  {Cognitive} {Strategies} in {Raven}’s {Advanced} {Progressive} {Matrices}:
  {A} {Data}-{Driven} {Perspective}.
\newblock Journal of Intelligence.  2025 Feb; 13(2):14.
\newblock \urlprefix\url{https://www.mdpi.com/2079-3200/13/2/14},
  \href{10.3390/jintelligence13020014}{\doiprefix
  \detokenize{10.3390/jintelligence13020014}}.

\bibitem[{Marjieh et~al.(2024)Marjieh, Raja and Sucholutsky, Ilia and van Rijn,
  Pol and Jacoby, Nori and Griffiths, Thomas L.}]{marjieh_large_2024}
\textbf{\color{eLifeMediumGrey} Marjieh R}, Sucholutsky I, van Rijn P, Jacoby
  N, Griffiths TL.
\newblock Large language models predict human sensory judgments across six
  modalities.
\newblock Scientific Reports.  2024 Sep; 14(1):21445.
\newblock \urlprefix\url{https://www.nature.com/articles/s41598-024-72071-1},
  \href{10.1038/s41598-024-72071-1}{\doiprefix
  \detokenize{10.1038/s41598-024-72071-1}}, publisher: Nature Publishing Group.

\bibitem[{Mischler et~al.(2024)Mischler, Gavin and Li, Yinghao Aaron and
  Bickel, Stephan and Mehta, Ashesh D. and Mesgarani,
  Nima}]{mischler_contextual_2024}
\textbf{\color{eLifeMediumGrey} Mischler G}, Li YA, Bickel S, Mehta AD,
  Mesgarani N.
\newblock Contextual {Feature} {Extraction} {Hierarchies} {Converge} in {Large}
  {Language} {Models} and the {Brain}; 2024,
  \urlprefix\url{https://arxiv.org/abs/2401.17671v1},
  \href{10.48550/arXiv.2401.17671}{\doiprefix
  \detokenize{10.48550/arXiv.2401.17671}}, arXiv:2401.17671 [cs.CL], version 1.

\bibitem[{Musker et~al.(2025)Musker, Sam and Duchnowski, Alex and Millière,
  Raphaël and Pavlick, Ellie}]{musker_llms_2025}
\textbf{\color{eLifeMediumGrey} Musker S}, Duchnowski A, Millière R, Pavlick
  E.
\newblock {LLMs} as models for analogical reasoning; 2025,
  \urlprefix\url{http://arxiv.org/abs/2406.13803},
  \href{10.48550/arXiv.2406.13803}{\doiprefix
  \detokenize{10.48550/arXiv.2406.13803}}, arXiv:2406.13803 [cs] version: 2.

\bibitem[{Newell and Simon(1956)Newell, A. and Simon, H.}]{newell_logic_1956}
\textbf{\color{eLifeMediumGrey} Newell A}, Simon H.
\newblock The logic theory machine–a complex information processing system.
\newblock IRE Transactions on Information Theory.  1956 Sep; 2(3):61--79.
\newblock \urlprefix\url{https://ieeexplore.ieee.org/document/1056797},
  \href{10.1109/TIT.1956.1056797}{\doiprefix
  \detokenize{10.1109/TIT.1956.1056797}}.

\bibitem[{Newell(1955)Newell, Allen}]{newell_chess_1955}
\textbf{\color{eLifeMediumGrey} Newell A}.
\newblock The chess machine: an example of dealing with a complex task by
  adaptation.
\newblock In: \emph{Proceedings of the {March} 1-3, 1955, {Western} {Joint}
  {Computer} {Conference}} {AFIPS} '55 ({Western}), New York, NY, USA:
  Association for Computing Machinery; 1955. p. 101--108.
\newblock \urlprefix\url{https://dl.acm.org/doi/10.1145/1455292.1455312},
  \href{10.1145/1455292.1455312}{\doiprefix
  \detokenize{10.1145/1455292.1455312}}.

\bibitem[{O'Connell et~al.(2012)O'Connell, Redmond G. and Dockree, Paul M. and
  Kelly, Simon P.}]{oconnell_supramodal_2012}
\textbf{\color{eLifeMediumGrey} O'Connell RG}, Dockree PM, Kelly SP.
\newblock A supramodal accumulation-to-bound signal that determines perceptual
  decisions in humans.
\newblock Nature Neuroscience.  2012; 15(12):1729--1735.
\newblock \href{10.1038/nn.3248}{\doiprefix \detokenize{10.1038/nn.3248}}.

\bibitem[{Perfetti et~al.(2009)Perfetti, Bernardo and Saggino, Aristide and
  Ferretti, Antonio and Caulo, Massimo and Romani, Gian Luca and Onofrj,
  Marco}]{perfetti_differential_2007}
\textbf{\color{eLifeMediumGrey} Perfetti B}, Saggino A, Ferretti A, Caulo M,
  Romani GL, Onofrj M.
\newblock Differential patterns of cortical activation as a function of fluid
  reasoning complexity.
\newblock Human Brain Mapping.  2009 Feb; 30(2):497--510.
\newblock
  \urlprefix\url{https://www.ncbi.nlm.nih.gov/pmc/articles/PMC6871137/},
  \href{10.1002/hbm.20519}{\doiprefix \detokenize{10.1002/hbm.20519}}.

\bibitem[{Pernet et~al.(2019)Pernet, Cyril R. and Appelhoff, Stefan and
  Gorgolewski, Krzysztof J. and Flandin, Guillaume and Phillips, Christophe and
  Delorme, Arnaud and Oostenveld, Robert}]{pernet_eeg-bids_2019}
\textbf{\color{eLifeMediumGrey} Pernet CR}, Appelhoff S, Gorgolewski KJ,
  Flandin G, Phillips C, Delorme A, Oostenveld R.
\newblock {EEG}-{BIDS}, an extension to the brain imaging data structure for
  electroencephalography.
\newblock Scientific Data.  2019 Jun; 6(1):103.
\newblock \urlprefix\url{https://www.nature.com/articles/s41597-019-0104-8},
  \href{10.1038/s41597-019-0104-8}{\doiprefix
  \detokenize{10.1038/s41597-019-0104-8}}.

\bibitem[{Perrin et~al.(1989)Perrin, F. and Pernier, J. and Bertrand, O. and
  Echallier, J. F.}]{perrin_spherical_1989}
\textbf{\color{eLifeMediumGrey} Perrin F}, Pernier J, Bertrand O, Echallier JF.
\newblock Spherical splines for scalp potential and current density mapping.
\newblock Electroencephalography and Clinical Neurophysiology.  1989 Feb;
  72(2):184--187.
\newblock
  \urlprefix\url{https://www.sciencedirect.com/science/article/pii/0013469489901806},
  \href{10.1016/0013-4694(89)90180-6}{\doiprefix
  \detokenize{10.1016/0013-4694(89)90180-6}}.

\bibitem[{Pl{\"o}chl et~al.(2012)Pl{\"o}chl, Michael and Ossand{\'o}n, Jos{\'e}
  P. and K{\"o}nig, Peter}]{plochl_combining_2012}
\textbf{\color{eLifeMediumGrey} Pl{\"o}chl M}, Ossand{\'o}n JP, K{\"o}nig P.
\newblock Combining {EEG} and eye tracking: identification, characterization,
  and correction of eye movement artifacts in electroencephalographic data.
\newblock Frontiers in Human Neuroscience.  2012; 6:278.
\newblock \href{10.3389/fnhum.2012.00278}{\doiprefix
  \detokenize{10.3389/fnhum.2012.00278}}.

\bibitem[{Reverberi et~al.(2005)Reverberi, Carlo and D’Agostini, Serena and
  Skrap, Miran and Shallice, Tim}]{reverberi_generation_2005}
\textbf{\color{eLifeMediumGrey} Reverberi C}, D’Agostini S, Skrap M, Shallice
  T.
\newblock Generation and recognition of abstract rules in different frontal
  lobe subgroups.
\newblock Neuropsychologia.  2005 Jan; 43(13):1924--1937.
\newblock
  \urlprefix\url{https://www.sciencedirect.com/science/article/pii/S0028393205001284},
  \href{10.1016/j.neuropsychologia.2005.03.004}{\doiprefix
  \detokenize{10.1016/j.neuropsychologia.2005.03.004}}.

\bibitem[{{rsatoolbox developers}(n.d.)}]{rsatoolbox_documentation_nd}
\textbf{\color{eLifeMediumGrey} {rsatoolbox developers}}, Python
  Representational Similarity Analysis toolbox ({rsatoolbox}): documentation;
  n.d.
\newblock \urlprefix\url{https://rsatoolbox.readthedocs.io/en/stable/},
  accessed 1 October 2026.

\bibitem[{Sajadieh et~al.(2026)Sajadieh, Sha and Fattorini, Loredana and
  Perrault, Raymond and Gil, Yolanda and Parli, Vanessa and Santarlasci, Lapo
  and Pava, Juan and Maslej, Nestor and Altman, Russ and Brynjolfsson, Erik and
  Brodley, Carla and Clark, Jack and Dignum, Virginia and Kumar, Vipin and
  Landay, James and Lyons, Terah and Manyika, James and Niebles, Juan Carlos
  and Shoham, Yoav and Tabassi, Elham and Wald, Russell and Walsh, Toby and
  Weld, Dan}]{sajadieh_ai_index_2026}
\textbf{\color{eLifeMediumGrey} Sajadieh S}, Fattorini L, Perrault R, Gil Y,
  Parli V, Santarlasci L, Pava J, Maslej N, Altman R, Brynjolfsson E, Brodley
  C, Clark J, Dignum V, Kumar V, Landay J, Lyons T, Manyika J, Niebles JC,
  Shoham Y, Tabassi E, et~al.
\newblock Artificial Intelligence Index Report 2026.
\newblock Stanford Institute for Human-Centered Artificial Intelligence; 2026.

\bibitem[{Santarnecchi et~al.(2017)Santarnecchi, Emiliano and Emmendorfer,
  Alexandra and Pascual-Leone, Alvaro}]{santarnecchi_dissecting_2017}
\textbf{\color{eLifeMediumGrey} Santarnecchi E}, Emmendorfer A, Pascual-Leone
  A.
\newblock Dissecting the parieto-frontal correlates of fluid intelligence: a
  comprehensive {ALE} meta-analysis study.
\newblock Intelligence.  2017 Jul; 63:9--28.
\newblock
  \urlprefix\url{https://www.sciencedirect.com/science/article/pii/S0160289617300090},
  \href{10.1016/j.intell.2017.04.008}{\doiprefix
  \detokenize{10.1016/j.intell.2017.04.008}}.

\bibitem[{Schrimpf et~al.(2021)Schrimpf, Martin and Blank, Idan Asher and
  Tuckute, Greta and Kauf, Carina and Hosseini, Eghbal A. and Kanwisher, Nancy
  and Tenenbaum, Joshua B. and Fedorenko, Evelina}]{schrimpf_neural_2021}
\textbf{\color{eLifeMediumGrey} Schrimpf M}, Blank IA, Tuckute G, Kauf C,
  Hosseini EA, Kanwisher N, Tenenbaum JB, Fedorenko E.
\newblock The neural architecture of language: integrative modeling converges
  on predictive processing.
\newblock Proceedings of the National Academy of Sciences of the United States
  of America.  2021 Nov; 118(45):e2105646118.
\newblock \href{10.1073/pnas.2105646118}{\doiprefix
  \detokenize{10.1073/pnas.2105646118}}.

\bibitem[{Silver et~al.(2016)Silver, David and Huang, Aja and Maddison, Chris
  J. and Guez, Arthur and Sifre, Laurent and van den Driessche, George and
  Schrittwieser, Julian and Antonoglou, Ioannis and Panneershelvam, Veda and
  Lanctot, Marc and Dieleman, Sander and Grewe, Dominik and Nham, John and
  Kalchbrenner, Nal and Sutskever, Ilya and Lillicrap, Timothy and Leach,
  Madeleine and Kavukcuoglu, Koray and Graepel, Thore and Hassabis,
  Demis}]{silver_mastering_2016}
\textbf{\color{eLifeMediumGrey} Silver D}, Huang A, Maddison CJ, Guez A, Sifre
  L, van~den Driessche G, Schrittwieser J, Antonoglou I, Panneershelvam V,
  Lanctot M, Dieleman S, Grewe D, Nham J, Kalchbrenner N, Sutskever I,
  Lillicrap T, Leach M, Kavukcuoglu K, Graepel T, Hassabis D.
\newblock Mastering the game of go with deep neural networks and tree search.
\newblock Nature.  2016 Jan; 529(7587):484--489.
\newblock \urlprefix\url{https://www.nature.com/articles/nature16961},
  \href{10.1038/nature16961}{\doiprefix \detokenize{10.1038/nature16961}},
  publisher: Nature Publishing Group.

\bibitem[{Silver et~al.(2018)Silver, David and Hubert, Thomas and
  Schrittwieser, Julian and Antonoglou, Ioannis and Lai, Matthew and Guez,
  Arthur and Lanctot, Marc and Sifre, Laurent and Kumaran, Dharshan and
  Graepel, Thore and Lillicrap, Timothy and Simonyan, Karen and Hassabis,
  Demis}]{silver_general_2018}
\textbf{\color{eLifeMediumGrey} Silver D}, Hubert T, Schrittwieser J,
  Antonoglou I, Lai M, Guez A, Lanctot M, Sifre L, Kumaran D, Graepel T,
  Lillicrap T, Simonyan K, Hassabis D.
\newblock A general reinforcement learning algorithm that masters chess, shogi,
  and go through self-play.
\newblock Science.  2018 Dec; 362(6419):1140--1144.
\newblock \urlprefix\url{https://www.science.org/doi/10.1126/science.aar6404},
  \href{10.1126/science.aar6404}{\doiprefix
  \detokenize{10.1126/science.aar6404}}, publisher: American Association for
  the Advancement of Science.

\bibitem[{Silver et~al.(2017)Silver, David and Schrittwieser, Julian and
  Simonyan, Karen and Antonoglou, Ioannis and Huang, Aja and Guez, Arthur and
  Hubert, Thomas and Baker, Lucas and Lai, Matthew and Bolton, Adrian and Chen,
  Yutian and Lillicrap, Timothy and Hui, Fan and Sifre, Laurent and van den
  Driessche, George and Graepel, Thore and Hassabis,
  Demis}]{silver_mastering_2017}
\textbf{\color{eLifeMediumGrey} Silver D}, Schrittwieser J, Simonyan K,
  Antonoglou I, Huang A, Guez A, Hubert T, Baker L, Lai M, Bolton A, Chen Y,
  Lillicrap T, Hui F, Sifre L, van~den Driessche G, Graepel T, Hassabis D.
\newblock Mastering the game of go without human knowledge.
\newblock Nature.  2017 Oct; 550(7676):354--359.
\newblock \urlprefix\url{https://www.nature.com/articles/nature24270},
  \href{10.1038/nature24270}{\doiprefix \detokenize{10.1038/nature24270}},
  publisher: Nature Publishing Group.

\bibitem[{{SR Research}(n.d.)}]{srresearch_eyelink1000plus_nd}
\textbf{\color{eLifeMediumGrey} {SR Research}}, {EyeLink 1000 Plus}; n.d.
\newblock \urlprefix\url{https://www.sr-research.com/eyelink-1000-plus/},
  accessed 1 October 2026.

\bibitem[{Tschentscher et~al.(2017)Tschentscher, Nadja and Mitchell, Daniel and
  Duncan, John}]{tschentscher_fluid_2017}
\textbf{\color{eLifeMediumGrey} Tschentscher N}, Mitchell D, Duncan J.
\newblock Fluid intelligence predicts novel rule implementation in a
  distributed frontoparietal control network.
\newblock Journal of Neuroscience.  2017 May; 37(18):4841--4847.
\newblock
  \urlprefix\url{https://www.ncbi.nlm.nih.gov/pmc/articles/PMC5426573/},
  \href{10.1523/JNEUROSCI.2478-16.2017}{\doiprefix
  \detokenize{10.1523/JNEUROSCI.2478-16.2017}}.

\bibitem[{Webb et~al.(2023)Webb, Taylor and Holyoak, Keith J. and Lu,
  Hongjing}]{webb_emergent_2023}
\textbf{\color{eLifeMediumGrey} Webb T}, Holyoak KJ, Lu H.
\newblock Emergent Analogical Reasoning in Large Language Models.
\newblock Nature Human Behaviour.  2023 Sep; 7(9):1526--1541.
\newblock \href{10.1038/s41562-023-01659-w}{\doiprefix
  \detokenize{10.1038/s41562-023-01659-w}}.

\bibitem[{Webb et~al.(2025)Webb, Taylor W and Holyoak, Keith J and Lu,
  Hongjing}]{webb_evidence_2025}
\textbf{\color{eLifeMediumGrey} Webb TW}, Holyoak KJ, Lu H.
\newblock Evidence from counterfactual tasks supports emergent analogical
  reasoning in large language models.
\newblock PNAS Nexus.  2025 May; 4(5):pgaf135.
\newblock \urlprefix\url{https://doi.org/10.1093/pnasnexus/pgaf135},
  \href{10.1093/pnasnexus/pgaf135}{\doiprefix
  \detokenize{10.1093/pnasnexus/pgaf135}}.

\bibitem[{Weindel et~al.(2025)Weindel, Gabriel and Borst, Jelmer P and van
  Maanen, Leendert}]{weindel_decision-making_2025}
\textbf{\color{eLifeMediumGrey} Weindel G}, Borst JP, van Maanen L.
\newblock Decision-making components and times revealed by the single-trial
  electroencephalogram.
\newblock eLife.  2025 Dec; 14:RP108049.
\newblock \urlprefix\url{https://doi.org/10.7554/eLife.108049},
  \href{10.7554/eLife.108049}{\doiprefix \detokenize{10.7554/eLife.108049}}.

\bibitem[{Weindel et~al.(2024)Weindel, Gabriel and van Maanen, Leendert and
  Borst, Jelmer P.}]{weindel_trial-by-trial_2024}
\textbf{\color{eLifeMediumGrey} Weindel G}, van Maanen L, Borst JP.
\newblock Trial-by-trial detection of cognitive events in neural time-series.
\newblock Imaging Neuroscience.  2024 Dec; 2:imag--2--00400.
\newblock \urlprefix\url{https://doi.org/10.1162/imag_a_00400},
  \href{10.1162/imag_a_00400}{\doiprefix \detokenize{10.1162/imag_a_00400}}.

\bibitem[{Wolf et~al.(2020)Wolf, Thomas and Debut, Lysandre and Sanh, Victor
  and Chaumond, Julien and Delangue, Clement and Moi, Anthony and Cistac,
  Pierric and Rault, Tim and Louf, Remi and Funtowicz, Morgan and Davison, Joe
  and Shleifer, Sam and von Platen, Patrick and Ma, Clara and Jernite, Yacine
  and Plu, Julien and Xu, Canwen and Le Scao, Teven and Gugger, Sylvain and
  Drame, Mariama and Lhoest, Quentin and Rush,
  Alexander}]{wolf_transformers_2020}
\textbf{\color{eLifeMediumGrey} Wolf T}, Debut L, Sanh V, Chaumond J, Delangue
  C, Moi A, Cistac P, Rault T, Louf R, Funtowicz M, Davison J, Shleifer S, von
  Platen P, Ma C, Jernite Y, Plu J, Xu C, Le~Scao T, Gugger S, Drame M, et~al.
\newblock {Transformers}: State-of-the-Art Natural Language Processing.
\newblock In: Liu Q, Schlangen D, editors. \emph{Proceedings of the 2020
  Conference on Empirical Methods in Natural Language Processing: System
  Demonstrations} Online: Association for Computational Linguistics; 2020. p.
  38--45.
\newblock \urlprefix\url{https://aclanthology.org/2020.emnlp-demos.6/},
  \href{10.18653/v1/2020.emnlp-demos.6}{\doiprefix
  \detokenize{10.18653/v1/2020.emnlp-demos.6}}.

\bibitem[{Xu et~al.(2025)Xu, Fengli and Hao, Qianyue and Shao, Chenyang and
  Zong, Zefang and Li, Yu and Wang, Jingwei and Zhang, Yunke and Wang, Jingyi
  and Lan, Xiaochong and Gong, Jiahui and Ouyang, Tianjian and Meng, Fanjin and
  Yan, Yuwei and Yang, Qinglong and Song, Yiwen and Ren, Sijian and Hu, Xinyuan
  and Feng, Jie and Gao, Chen and Li, Yong}]{xu_large_2025}
\textbf{\color{eLifeMediumGrey} Xu F}, Hao Q, Shao C, Zong Z, Li Y, Wang J,
  Zhang Y, Wang J, Lan X, Gong J, Ouyang T, Meng F, Yan Y, Yang Q, Song Y, Ren
  S, Hu X, Feng J, Gao C, Li Y.
\newblock Toward Large Reasoning Models: {{A}} Survey of Reinforced Reasoning
  with Large Language Models.
\newblock Patterns.  2025; 6(10):101370.
\newblock
  \urlprefix\url{https://www.cell.com/patterns/abstract/S2666-3899(25)00218-1},
  \href{10.1016/j.patter.2025.101370}{\doiprefix
  \detokenize{10.1016/j.patter.2025.101370}}.

\bibitem[{Yang et~al.(2025)Yang, Yukang and Campbell, Declan Iain and Huang,
  Kaixuan and Wang, Mengdi and Cohen, Jonathan D. and Webb, Taylor
  Whittington}]{yang_emergent_2025}
\textbf{\color{eLifeMediumGrey} Yang Y}, Campbell DI, Huang K, Wang M, Cohen
  JD, Webb TW.
\newblock Emergent Symbolic Mechanisms Support Abstract Reasoning in Large
  Language Models.
\newblock In: \emph{Proceedings of the 42nd International Conference on Machine
  Learning}, vol. 267 of Proceedings of Machine Learning Research PMLR; 2025.
  p. 70515--70549.
\newblock \urlprefix\url{https://proceedings.mlr.press/v267/yang25c.html}.

\bibitem[{Zurrin et~al.(2024)Zurrin, Riley and Wong, Samantha Tze Sum and Roes,
  Meighen M. and Percival, Chantal M. and Chinchani, Abhijit and Arreaza, Leo
  and Kusi, Mavis and Momeni, Ava and Rasheed, Maiya and Mo, Zhaoyi and
  Goghari, Vina M. and Woodward, Todd S.}]{zurrin_functional_2024}
\textbf{\color{eLifeMediumGrey} Zurrin R}, Wong STS, Roes MM, Percival CM,
  Chinchani A, Arreaza L, Kusi M, Momeni A, Rasheed M, Mo Z, Goghari VM,
  Woodward TS.
\newblock Functional brain networks involved in the raven's standard
  progressive matrices task and their relation to theories of fluid
  intelligence.
\newblock Intelligence.  2024 Mar; 103:101807.
\newblock
  \urlprefix\url{https://www.sciencedirect.com/science/article/pii/S0160289624000011},
  \href{10.1016/j.intell.2024.101807}{\doiprefix
  \detokenize{10.1016/j.intell.2024.101807}}.

\end{thebibliography}
\endgroup

\appendix
\newpage
\section{Supplementary results}
\label{sec:supp-detailed-results}
\captionsetup[table]{name=Table}
\subsection*{Human RDM construction and reliability}
\phantomsection\label{sec:supp-human-reliability}
RDM construction and equal weighting of sessions within participants and participants within the group are described in \hyperref[sec:rsa]{Methods, \textit{Representational similarity analysis}}; inclusion criteria are detailed in \hyperref[sec:sample-inclusion]{\textit{Sample inclusion}}. The reliability analyses used the condition-specific samples: \HierFRPParticipants{} participants and \HierFRPCompleteSessions{} sessions for FRPs, \HierResponseParticipants{} participants and \HierResponseSessions{} sessions for response-locked EEG, and \HierRestParticipants{} participants and \HierRestSessions{} sessions for pre-trial rest EEG.

Mean participant-to-leave-one-out-group (LOO) correlations and 95\% confidence intervals (CIs) supplement the main-text summary (\suppfigref{fig:participant-human-reliability}; procedure in \hyperref[meth:individual-reliability]{Methods, \textit{Individual-participant and reliability analyses}}). Across regions, mean similarity was $r=\HierReliabilityFRPMeanRSA{}$ (95\% CI [\HierReliabilityFRPCILow{}, \HierReliabilityFRPCIHigh{}]) for FRP, $r=\HierReliabilityResponseMeanRSA{}$ [\HierReliabilityResponseCILow{}, \HierReliabilityResponseCIHigh{}] for response, and $r=\HierReliabilityRestMeanRSA{}$ [\HierReliabilityRestCILow{}, \HierReliabilityRestCIHigh{}] for rest. Frontal means and 95\% CIs were $r=\HierReliabilityFrontalFRPMeanRSA{}$ [\HierReliabilityFrontalFRPCILow{}, \HierReliabilityFrontalFRPCIHigh{}], $r=\HierReliabilityFrontalResponseMeanRSA{}$ [\HierReliabilityFrontalResponseCILow{}, \HierReliabilityFrontalResponseCIHigh{}], and $r=\HierReliabilityFrontalRestMeanRSA{}$ [\HierReliabilityFrontalRestCILow{}, \HierReliabilityFrontalRestCIHigh{}], respectively.

\subsection*{Regional group-level LLM--human RSA}
\Cref{tab:supp-regional-rsa} provides the regional extrema underlying the group-level summaries in \Cref{fig:main-results}D. All entries use the matched sample of \HierMatchedParticipants{} participants and \HierMatchedCompleteSessions{} sessions. In particular, the low across-region rest mean conceals positive temporal and negative occipital correspondence; response-locked EEG also shows its highest mean in temporal channels.

\medskip
\noindent\begin{minipage}{\linewidth}
\centering
\captionof{table}{Group-level LLM--human RSA by EEG measure. Regional values are arithmetic means across the eight fixed LLMs; the across-region column averages the five regional means. Minima and maxima describe spatial variation, not uncertainty intervals.}
\label{tab:supp-regional-rsa}
\small
\setlength{\tabcolsep}{3pt}
\begin{tabular*}{\linewidth}{@{\extracolsep{\fill}}lrrll@{}}
\toprule
EEG measure & Across regions & Frontal & Minimum (region) & Maximum (region) \\
\midrule
FRP & \HierFRPMeanRSA{} & \HierFRPFrontalMeanRSA{} & \HierFRPRegionalRangeLow{} (temporal) & \HierFRPRegionalRangeHigh{} (frontal) \\
Response-locked & \HierResponseMeanRSA{} & \HierResponseFrontalMeanRSA{} & \HierResponseRegionalRangeLow{} (parietal) & \HierResponseRegionalRangeHigh{} (temporal) \\
Pre-trial rest & \HierRestMeanRSA{} & \HierRestFrontalMeanRSA{} & \HierRestRegionalRangeLow{} (occipital) & \HierRestRegionalRangeHigh{} (temporal) \\
\bottomrule
\end{tabular*}
\end{minipage}\par\medskip

\subsection*{Individual-participant LLM--human RSA}
\phantomsection\label{sec:supp-individual-rsa}
Complementing the frontal summaries in the main Results, mean individual-participant LLM RSA across the five regions was \HierIndividualFRPMeanRSA{} for FRP, \HierIndividualResponseMeanRSA{} for response, and \HierIndividualRestMeanRSA{} for rest. These analyses used the same condition-specific samples as the reliability analyses above.

The self-inclusive human reference correlates each participant's RDM with the full human group RDM, including that participant. Its values were \HierFullGroupFRPMeanRSA{}, \HierFullGroupResponseMeanRSA{}, and \HierFullGroupRestMeanRSA{} across regions, and \HierFullGroupFrontalFRPMeanRSA{}, \HierFullGroupFrontalResponseMeanRSA{}, and \HierFullGroupFrontalRestMeanRSA{} in frontal channels, for FRP, response, and rest, respectively. These values exceed the corresponding LOO references reported above, as expected when the participant contributes to the comparison RDM. They provide descriptive context for individual-participant alignment, rather than independent evidence of shared geometry.

\Needspace{22\baselineskip}
\subsection*{Exact pattern-identity alignment}
\phantomsection\label{sec:supp-pattern-tests}
\Cref{tab:supp-pattern-tests} reports all six exact one-sided pattern-label permutation tests, using the statistic and permutation procedure defined in \hyperref[meth:pattern-label-test]{Methods, \textit{Exact pattern-label permutation test}}. The frontal null distributions are shown in \suppfigref{fig:pattern-label-permutation}.

\medskip
\noindent\begin{minipage}{\linewidth}
\centering
\captionof{table}{Exact pattern-label permutation results for the matched sample. $T$ is the mean Fisher-transformed RSA across the eight LLMs and, for across-region tests, the five regions. Holm correction is applied separately within the three frontal tests and the three across-region tests. The unadjusted frontal FRP test evaluates the single primary hypothesis; its Holm-adjusted value is a supplementary robustness check.}
\label{tab:supp-pattern-tests}
\small
\setlength{\tabcolsep}{4pt}
\begin{tabular*}{\linewidth}{@{\extracolsep{\fill}}llrrr@{}}
\toprule
Regional scope & EEG measure & $T$ & Unadjusted $p$ & Holm-adjusted $p$ \\
\midrule
Frontal & FRP & \HierPatternLabelFrontalFRPObservedT{} & \HierPatternLabelFrontalFRPPGreater{} & \HierPatternLabelFrontalFRPPHolm{} \\
 & Response-locked & \HierPatternLabelFrontalResponseObservedT{} & \HierPatternLabelFrontalResponsePGreater{} & \HierPatternLabelFrontalResponsePHolm{} \\
 & Pre-trial rest & \HierPatternLabelFrontalRestObservedT{} & \HierPatternLabelFrontalRestPGreater{} & \HierPatternLabelFrontalRestPHolm{} \\
\midrule
Across regions & FRP & \HierPatternLabelOverallFRPObservedT{} & \HierPatternLabelOverallFRPPGreater{} & \HierPatternLabelOverallFRPPHolm{} \\
 & Response-locked & \HierPatternLabelOverallResponseObservedT{} & \HierPatternLabelOverallResponsePGreater{} & \HierPatternLabelOverallResponsePHolm{} \\
 & Pre-trial rest & \HierPatternLabelOverallRestObservedT{} & \HierPatternLabelOverallRestPGreater{} & \HierPatternLabelOverallRestPHolm{} \\
\bottomrule
\end{tabular*}
\end{minipage}\par\medskip

The primary frontal FRP result remained significant under the broader Holm correction across frontal measures. Neither frontal control nor any across-region test survived correction at $\alpha=.05$. These tests assess alignment within each EEG measure; they do not establish a difference in alignment between measures.

\clearpage
\section{Supplementary figures}
\captionsetup[figure]{name=Figure}

\noindent\begin{minipage}{\linewidth}
    \centering
    \includegraphics[width=1.0\linewidth]{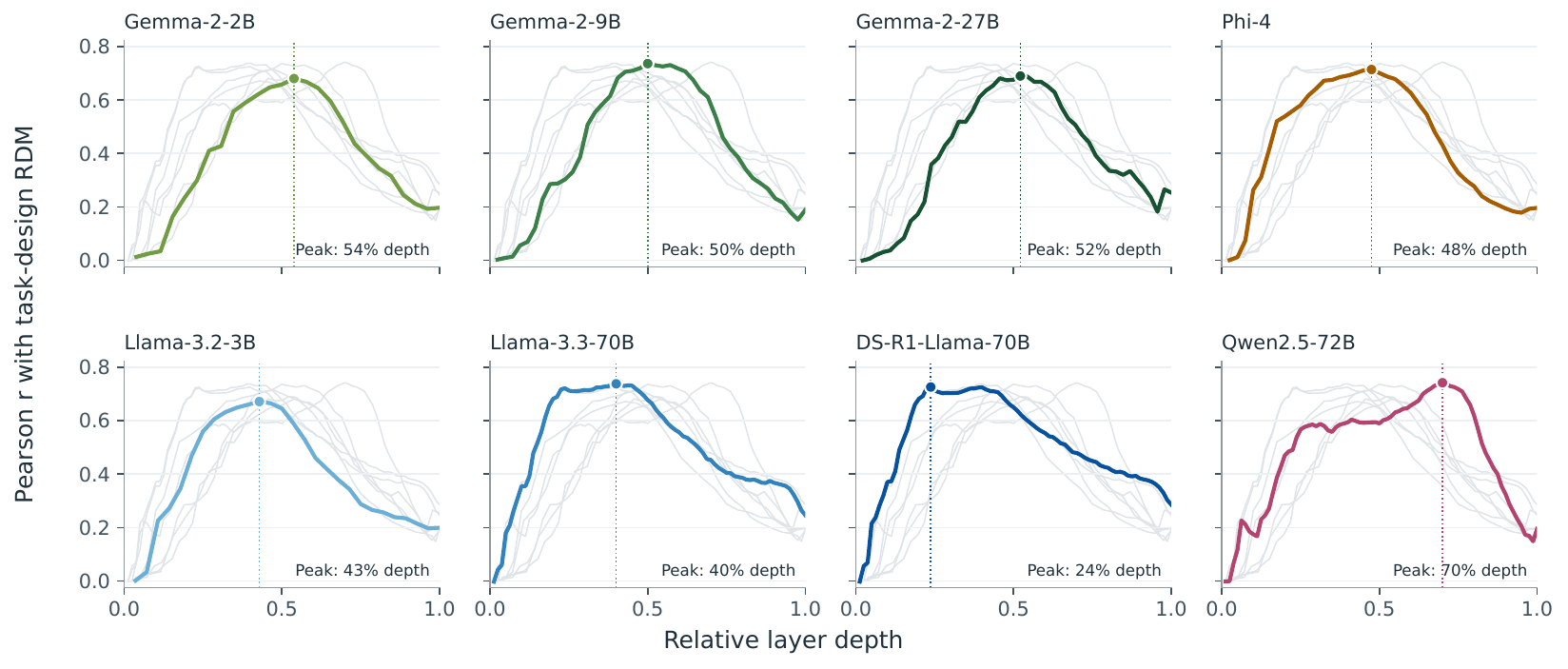}
    \captionof{figure}{Abstract-pattern structure across LLM depth. Each panel highlights one model's layer-wise Pearson correlation between its $400\times400$ trial RDM and the task-design RDM, which assigns zero dissimilarity to trials from the same abstract pattern and one to trials from different patterns. Colored curves show the named model; gray curves show the other seven models for context. The dot and vertical dotted line identify the maximum-correlation layer used for the primary LLM--human comparison, selected independently of EEG. Layers are numbered from one; relative depth is the layer number divided by the total number of layers; annotations give the selected depth as a percentage. All panels share the same axes.}
    \label{fig:llm-task-layer-profiles}
\end{minipage}\par

\clearpage

\begin{center}
    \includegraphics[width=\linewidth]{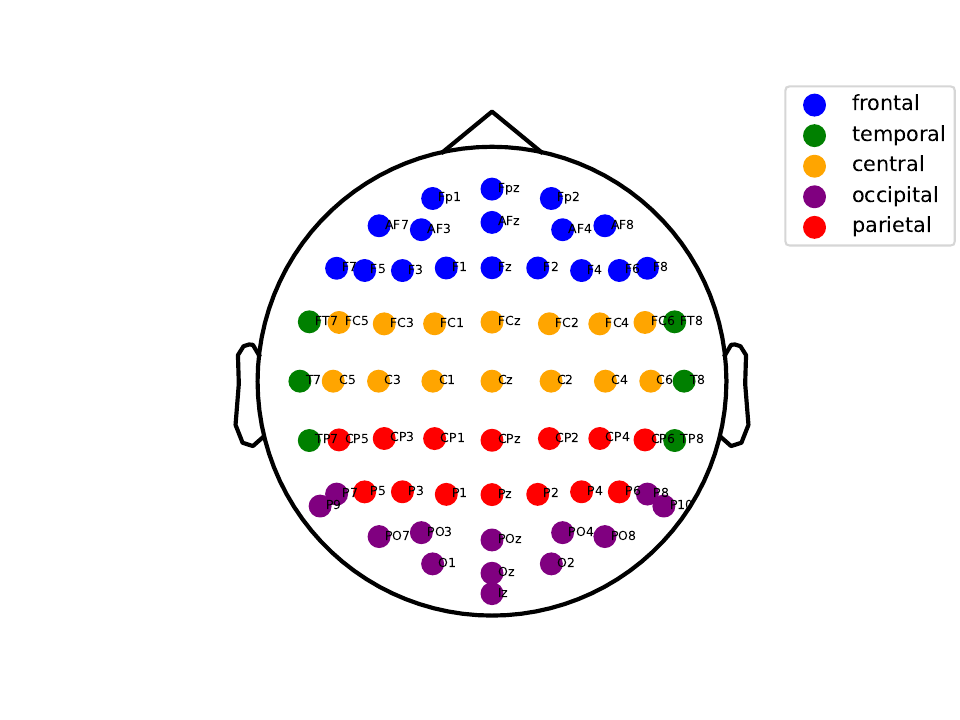}
    \captionof{figure}{EEG channel-group definitions for the full 64-electrode montage. The five primary regional groups are non-overlapping and together cover the montage; they share participants and are not independent statistical samples.}
    \label{fig:eeg-channel-groups}
\end{center}

\clearpage
\noindent\begin{minipage}{\linewidth}
    \centering
    \includegraphics[width=\linewidth]{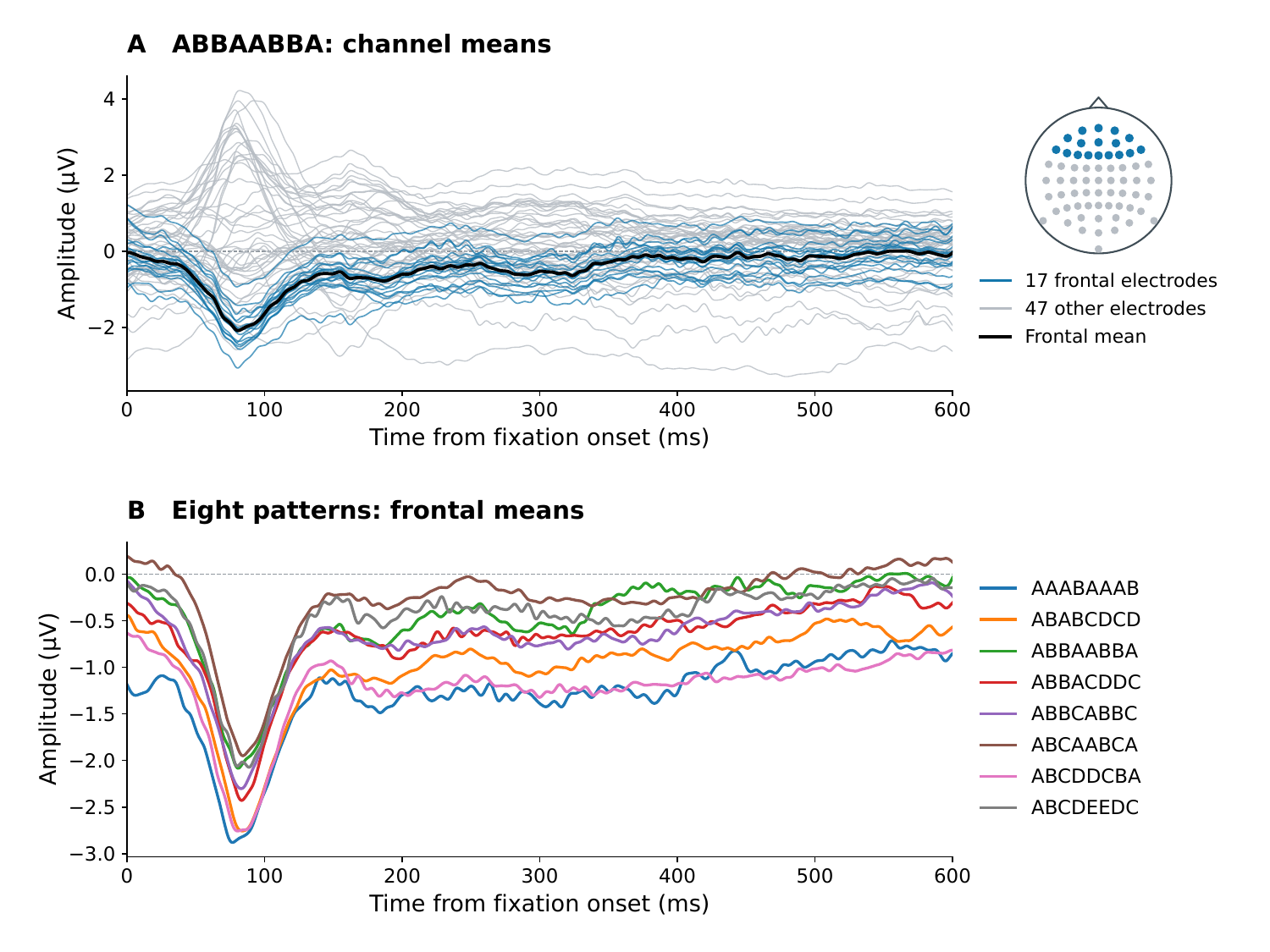}
    \captionsetup{width=\linewidth}
    \captionof{figure}{Fixation-related potential waveforms in the matched neural sample. \textbf{A:} Grand-average waveforms for the ABBAABBA pattern at all 64 electrodes; blue traces identify the 17 frontal electrodes, gray traces the remaining electrodes, and black their frontal mean. The inset shows electrode positions, nose up. \textbf{B:} Frontal mean waveforms for all eight abstract patterns. Time zero marks fixation onset. Trial FRPs were averaged within each pattern and session, then across sessions within each participant, and finally equally across \HierMatchedParticipants{} participants (\HierMatchedCompleteSessions{} sessions). The plots illustrate waveform morphology and do not constitute a statistical test of pattern differences.}
    \label{fig:matched-frp-waveforms}
\end{minipage}\par

\clearpage
\noindent\begin{minipage}{\linewidth}
    \centering
    \includegraphics[width=0.9\linewidth]{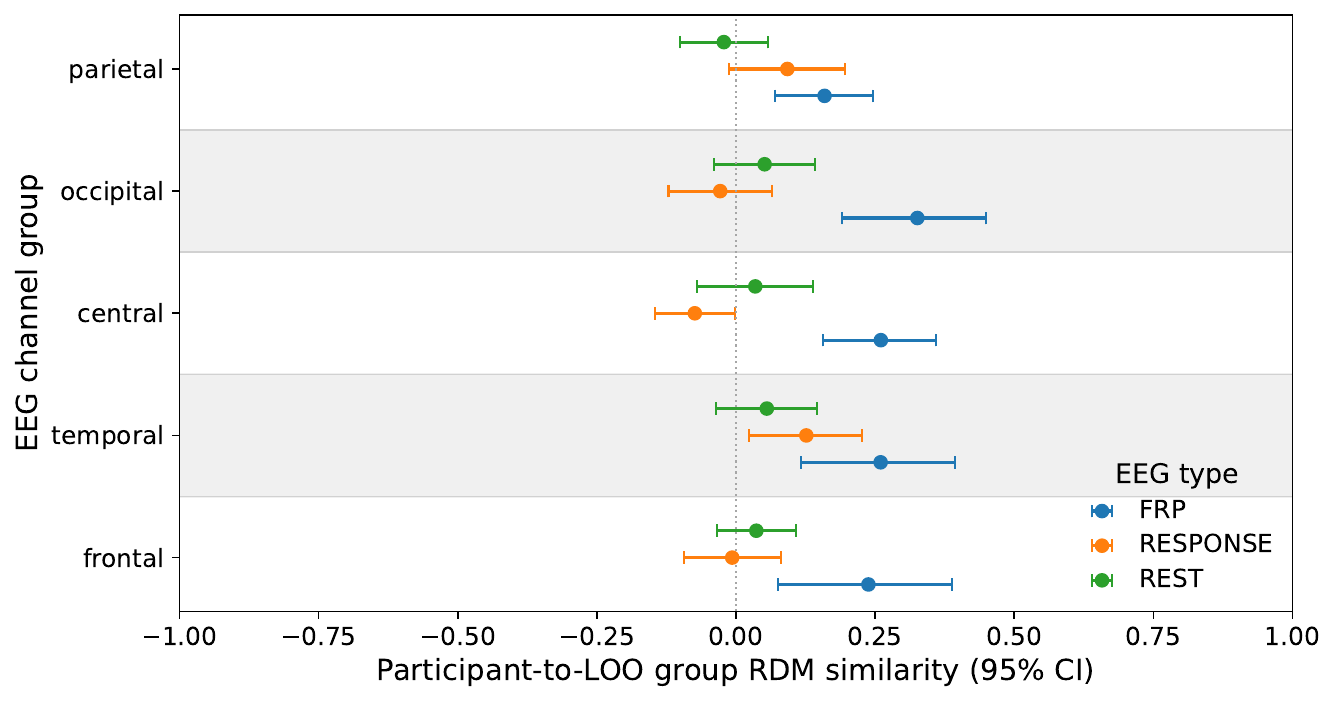}
    \captionsetup{width=\linewidth}
    \captionof{figure}{Participant-to-human RDM similarity for FRP, response-locked, and pre-trial rest EEG. Each point summarizes participant-to-leave-one-out-group Pearson correlations for one regional group and EEG measure: correlations were averaged in Fisher-$z$ space and back-transformed to $r$. Horizontal bars show 95\% Student-$t$ confidence intervals for the mean in Fisher-$z$ space, also back-transformed to $r$; colours distinguish EEG measures. FRP summaries use $N=\HierFRPParticipants{}$ participants with \HierFRPCompleteSessions{} complete sessions, whereas response and rest each use $N=\HierResponseParticipants{}$ participants with \HierResponseSessions{} sessions. These are group summaries, not individual participant points. The plot assesses shared human representational geometry rather than an LLM comparison.}
    \label{fig:participant-human-reliability}
\end{minipage}\par

\clearpage
\noindent\begin{minipage}{\linewidth}
    \centering
    \includegraphics[width=0.9\linewidth]{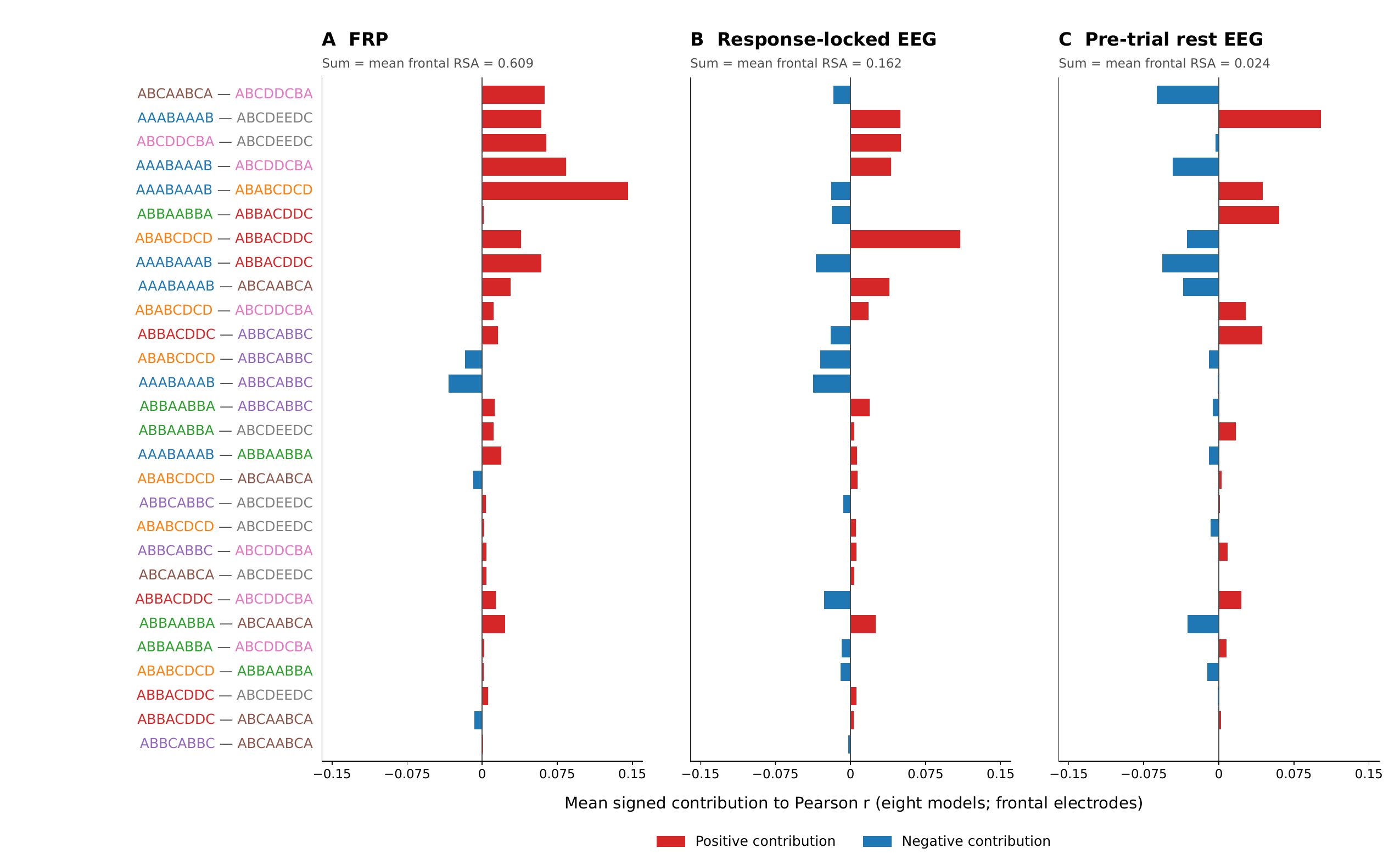}
    \captionsetup{width=\linewidth}
    \captionof{figure}{Pattern-pair contributions to frontal LLM--human RSA. Each bar shows one abstract-pattern pair's signed contribution to Pearson correlation, averaged across eight LLMs using the frontal human group RDM. Panels show \textbf{A} FRP, \textbf{B} response-locked EEG, and \textbf{C} pre-trial rest EEG for the matched sample of \HierMatchedParticipants{} participants and \HierMatchedCompleteSessions{} sessions. For each model and EEG measure, the 28 human and model RDM distances were separately centered and divided by their sample standard deviations. Each pattern pair's contribution was calculated as the product of its two standardized distances divided by 27 ($28-1$); these contributions were then averaged across the eight models. Consequently, summing all 28 bars in each panel recovers the mean frontal Pearson correlation printed above it. Red indicates positive contributions: a pair is relatively similar in both representations or relatively dissimilar in both. Blue indicates negative contributions: the relative similarity relationships are opposite. Bar colors reflect the sign of the contribution averaged across models, not necessarily agreement among every model. Colored pattern labels identify the two patterns in each pair. These descriptive contributions are not separate correlations or significance tests for individual pattern pairs.}
    \label{fig:matched-rdm-pair-contributions}
\end{minipage}\par

\bigskip
\noindent\begin{minipage}{\linewidth}
    \centering
    \includegraphics[width=0.9\linewidth]{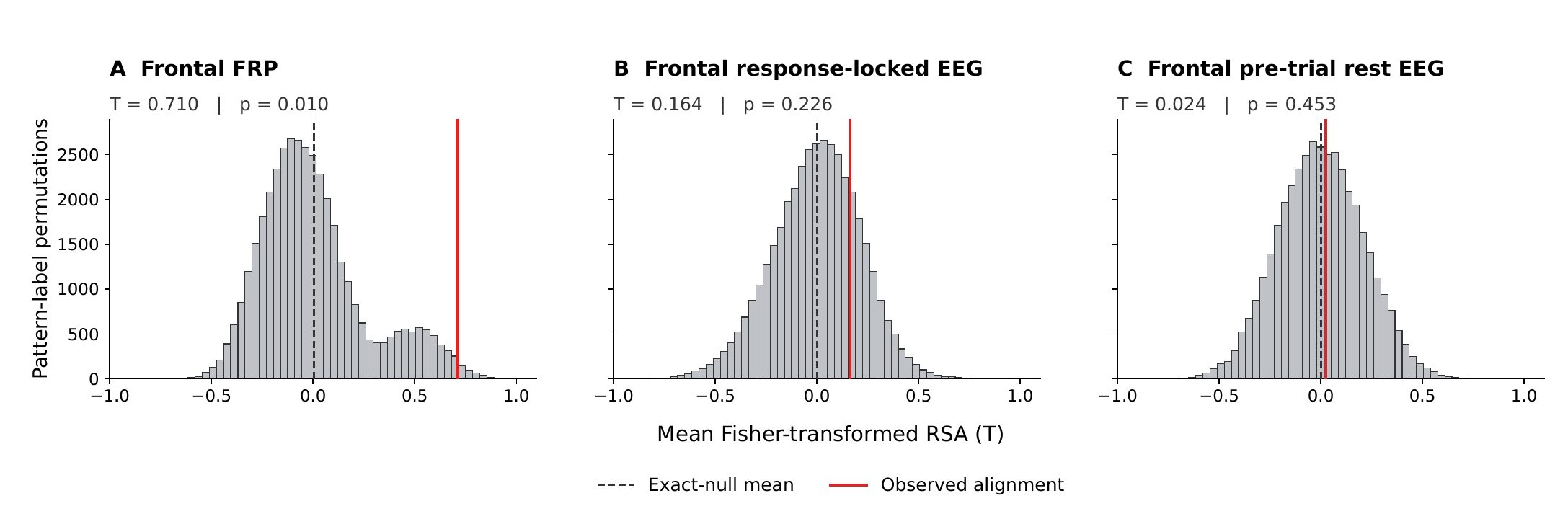}
    \captionsetup{width=\linewidth}
    \captionof{figure}{Exact pattern-label permutation tests of frontal LLM--human alignment. Histograms show the null distribution of $T$, the mean Fisher-transformed RSA across eight LLMs, for \textbf{A} frontal FRPs, \textbf{B} frontal response-locked EEG, and \textbf{C} frontal pre-trial rest EEG. All tests use the matched sample of \HierMatchedParticipants{} participants and \HierMatchedCompleteSessions{} sessions. Human RDM rows and corresponding columns were jointly relabelled under all $8!=40{,}320$ pattern permutations, while LLM RDMs remained fixed. Red lines mark observed alignment; dashed black lines mark the exact-null means. Displayed p-values are unadjusted, one-sided permutation probabilities of obtaining $T$ at least as large as observed. Frontal FRP alignment tests the primary hypothesis; response and rest provide control tests. These tests assess alignment within each EEG measure, not differences between measures.}
    \label{fig:pattern-label-permutation}
\end{minipage}\par

\clearpage
\noindent\begin{minipage}{\linewidth}
    \centering
    \includegraphics[width=\linewidth]{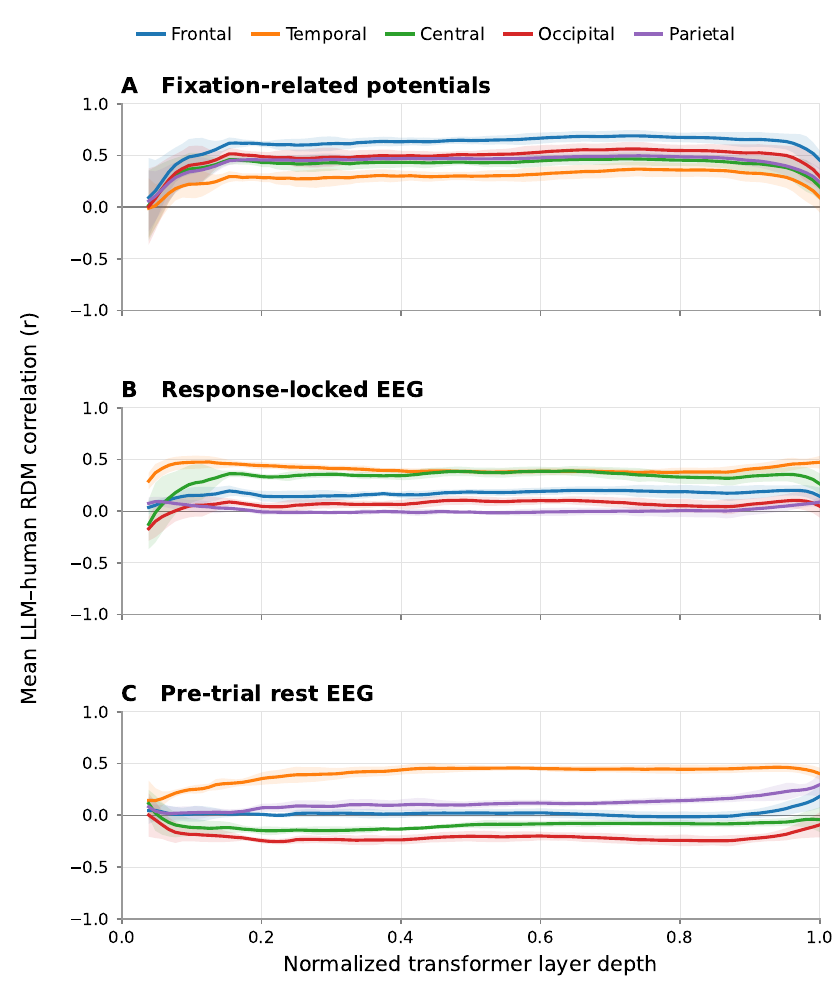}
    \captionsetup{width=\linewidth}
    \captionof{figure}{Mean layer-wise LLM--human RSA for \textbf{A} fixation-related potentials, \textbf{B} response-locked EEG, and \textbf{C} pre-trial rest EEG. All three conditions use the same \HierMatchedCompleteSessions{} sessions from \HierMatchedParticipants{} participants, with sessions averaged within participants and equal weight per participant. Each curve shows the mean Pearson correlation across eight LLMs for one electrode group after interpolation to a common normalized layer-depth grid. Shading spans $\pm1$ standard deviation across models. Electrode-group colors and axis limits are shared across panels. The profiles are descriptive because successive layers within a model are dependent.}
    \label{fig:supp-layerwise-eeg-regions}
\end{minipage}\par

\clearpage
\noindent\begin{minipage}{\linewidth}
    \centering
    \includegraphics[width=\linewidth]{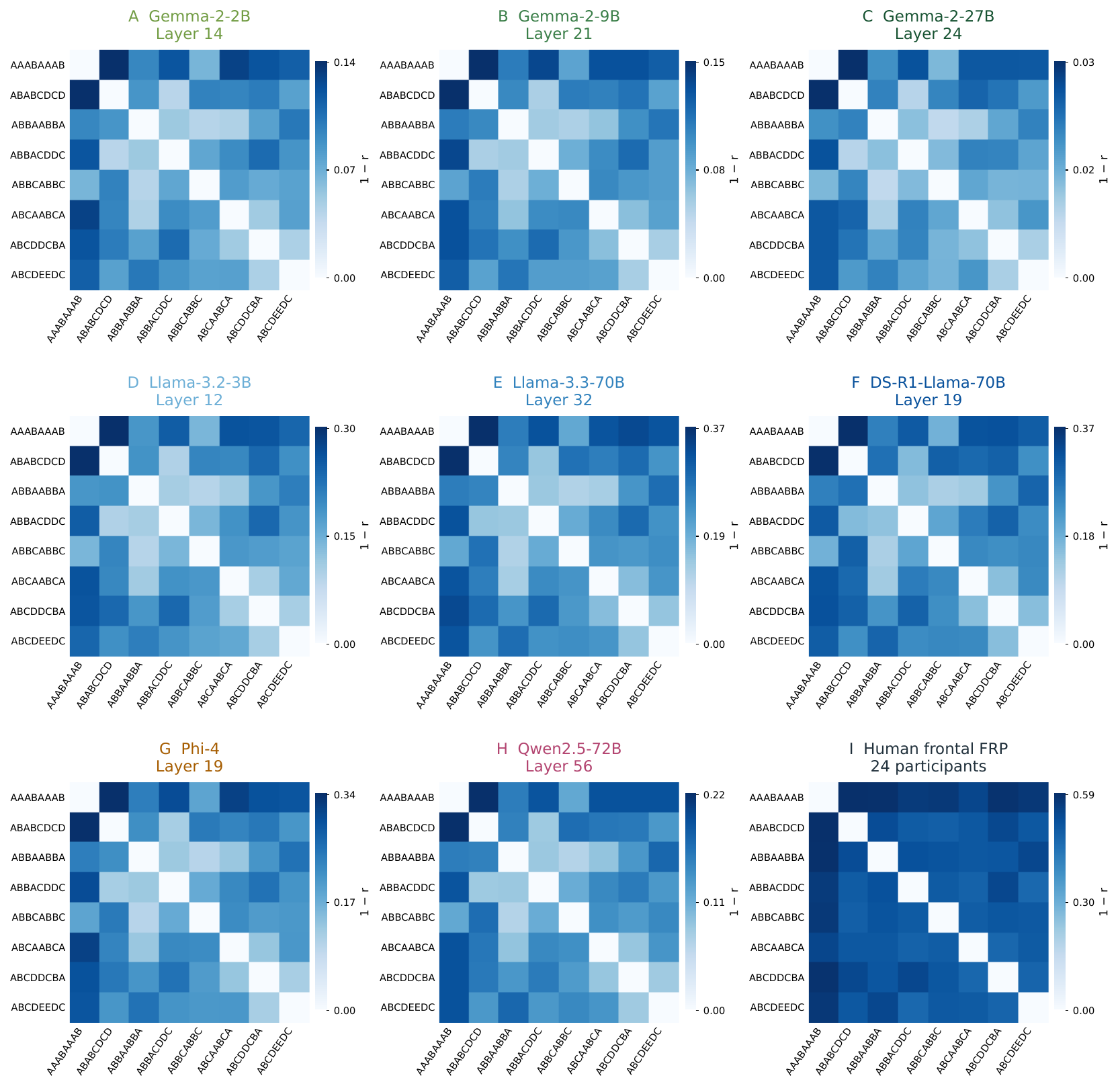}
    \captionsetup{width=\linewidth}
    \captionof{figure}{Task-optimal LLM RDMs and the human frontal-FRP group RDM. \textbf{A--H:} Pattern-level RDMs for each of the eight LLMs, computed from pattern-mean activation vectors at the layer maximizing trial-level similarity to the task-design RDM, independently of human EEG. Layer numbers start at one. \textbf{I:} Human frontal-FRP group RDM for the matched set of \HierMatchedParticipants{} participants and \HierMatchedCompleteSessions{} complete participant--session cells. Session RDMs were averaged within participant, then equally across participants. All panels show the same eight patterns in the same order; entries are correlation distances ($1-r$). Each color scale spans zero to that panel's maximum to reveal within-RDM structure; equal colors across panels do not imply equal distances. These are the pattern-level matrices used for the primary frontal LLM--human RSA.}
    \label{fig:supp-optimal-layer-rdms}
\end{minipage}\par

\end{document}